\documentclass[lettersize,journal]{IEEEtran}
\usepackage{amsmath,amsfonts}
\usepackage{algorithmic}
\usepackage{algorithm}
\usepackage{array}
\usepackage[caption=false,font=normalsize,labelfont=sf,textfont=sf]{subfig}
\usepackage{textcomp}
\usepackage{stfloats}
\usepackage{url}
\usepackage{verbatim}
\usepackage{graphicx}
\usepackage{cite}
\usepackage[colorlinks,linkcolor=blue,anchorcolor=blue,citecolor=blue]{hyperref}
\usepackage{textcomp}
\usepackage{multirow}
\usepackage{pifont}
\usepackage{mathrsfs}
\usepackage{booktabs}
\usepackage{makecell}

\usepackage{diagbox}
\usepackage{bbding}
\usepackage{textcomp}
\usepackage{xcolor}

\usepackage[switch]{lineno}
\modulolinenumbers[5]

\begin{document}
\title{DAWN: Domain-Adaptive Weakly Supervised Nuclei Segmentation via Cross-Task Interactions}

\author{Ye Zhang, Yifeng Wang, Zijie Fang, Hao Bian, Linghan Cai, Ziyue Wang, and\\ Yongbing Zhang, \IEEEmembership{Senior Member, IEEE}
\thanks{This work was supported in part by the National Natural Science Foundation of China under 62031023 \& 62331011; in part by the Shenzhen Science and Technology Project under GXWD20220818170353009, and in part by the Fundamental Research Funds for the Central Universities under No.HIT.OCEF.2023050. \textit{(Corresponding author: Yongbing Zhang.)}}

\thanks{Ye Zhang, Linghan Cai, Ziyue Wang, and Yongbing Zhang are with the School of Computer Science and Technology, Harbin Institute of Technology, Shenzhen, 518055, China. (e-mail: zhangye94@stu.hit.edu.cn; cailh@stu.hit.edu.cn; 200111326@stu.hit.edu.cn; ybzhang08@hit.edu.cn). Ye Zhang is also a joint PhD student at Leibniz-Institut für Analytische Wissenschaften – ISAS – e.V., Germany. (e-mail: ye.zhang@isas.de).}
\thanks{Yifeng Wang is with the School of Science, Harbin Institute of Technology, Shenzhen, 518055, China. (e-mail: wangyifeng@stu.hit.edu.cn).}
\thanks{Zijie Fang and Hao Bian are with the Tsinghua Shenzhen International Graduate School, Tsinghua University, Shenzhen, 518071, China. (e-mail: fzj22@mails.tsinghua.edu.cn; bianh21@mails.tsinghua.edu.cn).}

}

\markboth{Journal of \LaTeX\ Class Files,~Vol.~14, No.~8, March~2024}%
{Shell \MakeLowercase{\textit{et al.}}: A Sample Article Using IEEEtran.cls for IEEE Journals}

\maketitle

\begin{abstract}

Weakly supervised segmentation methods have garnered considerable attention due to their potential to alleviate the need for labor-intensive pixel-level annotations during model training. Traditional weakly supervised nuclei segmentation approaches typically involve a two-stage process: pseudo-label generation followed by network training. The performance of these methods is highly dependent on the quality of the generated pseudo-labels, which can limit their effectiveness. In this paper, we propose a novel domain-adaptive weakly supervised nuclei segmentation framework that addresses the challenge of pseudo-label generation through cross-task interaction strategies. Specifically, our approach leverages weakly annotated data to train an auxiliary detection task, which facilitates domain adaptation of the segmentation network. To improve the efficiency of domain adaptation, we introduce a consistent feature constraint module that integrates prior knowledge from the source domain. Additionally, we develop methods for pseudo-label optimization and interactive training to enhance domain transfer capabilities. We validate the effectiveness of our proposed method through extensive comparative and ablation experiments conducted on six datasets. The results demonstrate that our approach outperforms existing weakly supervised methods and achieves performance comparable to or exceeding that of fully supervised methods. Our code is available at \href{https://github.com/zhangye-zoe/DAWN}{https://github.com/zhangye-zoe/DAWN}.

\end{abstract}


\section{Introduction}
Nuclei segmentation plays a crucial role in computational pathology, serving as a fundamental step for various pathological diagnosis and patient prognosis tasks \cite{lee2013automatic, li2023erdunet, yun2023spectr, zhangboundary}. Meanwhile, accurate segmentation enables advanced nuclei analysis, including nuclei classification \cite{lagree2021review, bayramoglu2016transfer, doan2022sonnet}, nuclei counting \cite{graham2024conic, liu2022simultaneous}, and nuclei tracing \cite{ronneberger2015u, arbelle2018probabilistic, ulman2017objective}. However, developing fully supervised nuclei segmentation models for whole slide images (WSIs) poses challenges because of the extensive requirement of precise pixel-wise annotations in these gigapixel-sized images. To ease the manual annotation burden, the weakly supervised segmentation algorithm provides a promising solution. It leverages weakly supervised annotation information to excavate the morphology and structure of objects, enabling more efficient and accurate nucleic analysis.

\begin{figure}[t]
\centering
\includegraphics[width=3.3in]{"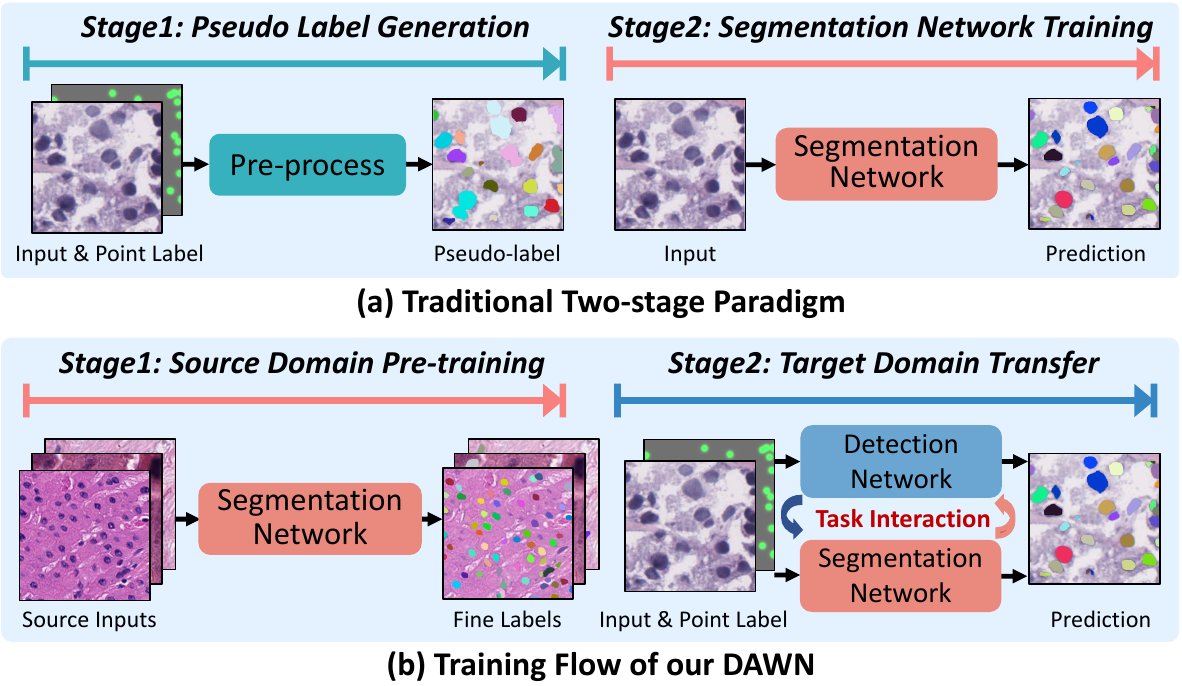"}
\caption{\textbf{The weakly supervised nuclei segmentation flows based on point annotation.} (a) represents a traditional two-stage segmentation paradigm. The point annotation is employed to generate the pseudo label, which is then used to train the segmentation network. (b) represents our domain-adaptive weakly supervised segmentation method. The point annotation is utilized to train a detection network, which can assist target domain transfer.} \label{fig: para}
\end{figure}

\begin{figure}[t]
\centering
\includegraphics[width=3.3in]{"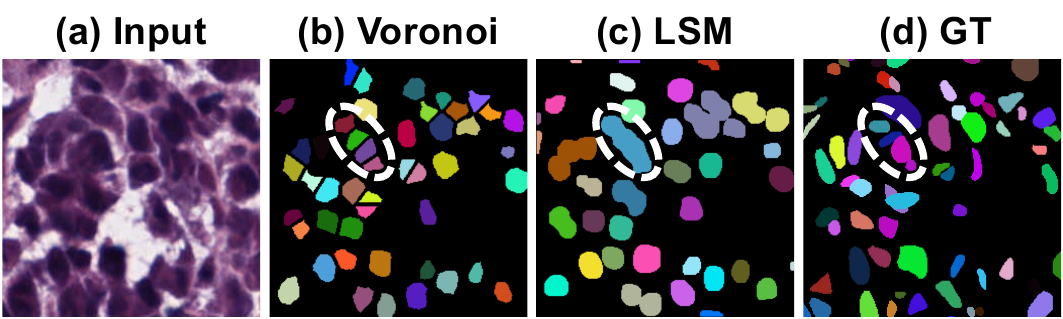"}
\caption{\textbf{The problems of generated pseudo labels.} (a) represents
the input image; (b) represents the generated mask using Voronoi \cite{qu2019weakly}; (c) represents the generated mask using LSM \cite{zhang2022ddtnet}; (d) is the ground-truth label of input.} \label{fig: point}
\end{figure}

Weakly supervised methods based on point annotation have recently been widely adopted for nuclei segmentation \cite{qu2019weakly,zhao2020weakly,tian2020weakly}, which reduce manual consumption by a large margin. Due to inadequate supervision information, these methods generally need a pre-processing step to generate masks. Then, these generated masks are utilized to guide the training of the segmentation network, as shown in \ref{fig: para} (a). However, the two-stage training flow faces two significant challenges. \textbf{On the one hand}, the generated pseudo-labels often are coarse and lack precise representation of nuclear morphology. For example, the Voronoi method \cite{qu2019weakly} utilizes distance measurements to differentiate overlapping nuclei, which results in the disruption of boundary morphology, as depicted in Fig. \ref{fig: point} (b). The level set method (LSM) \cite{zhang2022ddtnet} employs color features to simulate nuclear contours. But, it fails to distinguish cytoplasm and nucleus with similar colors, as illustrated in Fig. \ref{fig: point} (c). \textbf{On the other hand}, these two-stage methods still suffer from the influences of inaccurate pseudo-labels. Although some localized optimization methods have been proposed to enhance segmentation performance, such as boundary mining \cite{lin2024bonus} and conditional random fields (CRF) \cite{qu2020weakly}, these localized optimizing methods fail to adapt to the entire training process, resulting in segmentation outcomes that remain unsatisfactory. Therefore, it is necessary to adopt a new training paradigm to improve the nuclear segmentation performance in weakly supervised scenarios.

Recent research has highlighted the potential of domain adaptation (DA) for improving segmentation tasks. By utilizing task-relevant information from the source domain, DA can improve instance segmentation performance in the target domain. However, existing unsupervised domain adaptation methods \cite{chen2019synergistic, liu2020pdam, hsu2021darcnn} often struggle with dense segmentation tasks like nuclei segmentation, primarily due to their limited ability to incorporate target domain information effectively. In weakly supervised scenarios, point-annotated images provide valuable prior knowledge regarding the location and distribution of nuclei. This additional information can be leveraged to enhance domain adaptation models. To address this, we propose a novel weakly supervised nuclei segmentation network incorporating domain adaptation techniques. Our approach presents several advantages over previous models. \textbf{First}, unlike traditional weakly supervised segmentation methods that rely heavily on generating pseudo-labels for training \cite{piva2023empirical, shin2023sdc}, our method employs weak annotation to train auxiliary task to boost domain transfer capabilities. This approach minimizes the issues associated with inaccurate pseudo-labels. \textbf{Second}, our weakly supervised segmentation model provides distinct advantages over unsupervised methods by incorporating additional supervisory information. This integration allows us to leverage weak annotation information more effectively, significantly enhancing domain transfer capability \cite{shu2019transferable, baek2020weakly, inoue2018cross}.

Motivated by these observations, this paper proposes a \textbf{D}omain-\textbf{A}daptive \textbf{W}eakly supervised \textbf{N}uclei segmentation method (DAWN). By utilizing point annotation data, DAWN incorporates a cross-task interaction strategy at both the feature and task levels to enhance the model's domain transfer capability for accurate nuclei segmentation. 
Firstly, we propose a consistent feature constraint (CFC) module, which leverages the task correlation between the detection and segmentation networks to preserve prior knowledge related to the target data.
Secondly, we introduce a task-level supervision information optimization strategy to address the challenge of the domain mismatch problem. By leveraging the strong localization ability of the detection network, we compensate for the output shortcomings of the segmentation network.
Lastly, we devise an interactive training method to achieve progressive transfer of the segmentation network. In detail, we establish mutual supervision between the segmentation and detection tasks. This approach enhances the performance of segmentation predictions by leveraging the complementary aspects of both tasks.
Extensive experiment results demonstrate the superiority of the proposed framework. Overall, our contributions can be summarized as follows:

\begin{itemize}
\item  We propose a novel domain-adaptive weakly supervised nuclei segmentation framework that departs from traditional two-stage weakly supervised methods. Our approach leverages a detection task with point annotations and domain transfer strategies for accurate segmentation.
\item We devise a consistent feature constraint module to retain the prior information from the source domain, which boosts the domain transfer efficiency at the feature level. 
\item We develop a combined pseudo-label optimization method incorporating the outputs  from segmentation and detection networks. Besides, through an iterative training strategy, this module enhances the domain transfer capability at the task level.
\item We propose an interactive supervised training method that leverages optimized pseudo-labels to supervise the training of segmentation and detection networks simultaneously. This approach facilitates the learning of complementary features between the two networks.
\item We validate our method on six datasets. Extensive experiments indicate that our method outperforms existing weakly supervised methods and achieves equal or even better performance with fully supervised methods.
\end{itemize}

\section{Related Work}
\subsection{Nuclei Instance Segmentation}
Pathological image analysis based on nuclear segmentation \cite{zhou2018unet++, jin2019dunet,jin2020ra, ke2023clusterseg, yao2024position} has been widely discussed. Currently, some representative work mainly focuses on the problem of nuclear boundary overlapping. Such as DCAN \cite{chen2016dcan} adds the contour prediction task to the semantic segmentation model to separate instances. HoverNet\cite{graham2019hover} uses horizontal and vertical gradients to distinguish overlapping instances. Cellpose\cite{stringer2021cellpose} uses simulated diffusion and spatial gradients to trace the instances to which each pixel belongs. In addition, other methods \cite{he2021cdnet, jiang2023donet} are also proposed to solve the boundary overlapping problem.

Although fully supervised nuclei segmentation has achieved good progress, it is expensive and labor intensive because it requires a lot of manual annotation at the pixel level. In contrast, weakly supervised methods provide an effective solution to reduce the reliance of model development on annotations. Mainstream weakly supervised nuclear segmentation models follow a two-stage training paradigm. First, point annotation images are used to generate pseudo-labels, and then these generated pseudo-labels are used to supervise network training \cite{qu2019weakly,dong2020towards,guo2021learning, zhang2023weakly}. However, these two-stage training methods are affected by initialized pseudo-labels, and there is no explicit optimization in the training process, so the model performance is poor. In this paper, we propose a domain adaptation method for weakly supervised nuclear segmentation. Our method uses end-to-end training to avoid the traditional two-stage training paradigm and significantly improves the accuracy of nuclear segmentation in weakly supervised scenarios.

\subsection{Weakly Supervised Segmentation Methods}
Existing weakly supervised segmentation approaches utilize various types of partially labeled annotation to help automatically pixel-level segmentation. These weak labels include image-level, box-level, and point labels. Among these, methods relying on image-level annotations typically use class activation maps (CAMs) \cite{selvaraju2016grad} to generate pseudo-labels. For instance, PRM \cite{zhou2018weakly} employs CAM peak backpropagation to identify highly informative object regions. SC-CAM \cite{chang2020weakly} enhances CAM completeness by incorporating information from unsupervised clustering. RRM \cite{zhang2020reliability} combines CAM with a reliable score evaluation to produce more refined pseudo-labels. In addition, several other models \cite{zhang2018spftn, xu2024multidimensional} are proposed to solve weakly supervised video segmentation tasks. The box-level annotations are often used in conjunction with saliency models. BoxCaseg \cite{wang2021weakly} leverages a salient segmentation head to improve object localization, while BoxTeacher \cite{cheng2023boxteacher} applies noise-aware pixel loss and noise-reduced affinity loss to optimize pseudo-labels. Point labels \cite{han2019weakly, yao2024position, wang2024dynamic} are widely used in small object segmentation tasks because of their excellent location ability. ASDT \cite{zhang2023weakly} propagates labels between image and pixel levels, thereby improving semantic segmentation performance.
Despite the success of existing weakly supervised segmentation methods, these approaches often struggle with accurate shape prediction due to insufficient consideration of the morphological characteristics of instances. In this paper, we introduce the domain adaptive method into the weakly supervised segmentation model and learn morphological features with the prior knowledge of external data, thus improving the segmentation performance.

\subsection{Domain Adaptation Segmentation}

Pre-training methods \cite{kirillov2023segment} have made it possible to train models with limited samples. However, directly applying pre-trained models to target domains often fails due to distribution gaps between source and target samples. This challenge has led to the development of domain adaptation techniques. For example, Wang et al. \cite{he2021multi} use image translation methods to align pixel value distributions, thereby reducing the domain gap. Similarly, DIGA \cite{wang2023dynamically} introduces both distribution adaptation and semantic adaptation modules to facilitate joint model adaptation.
ADPL \cite{cheng2023adpl} improves domain adaptation by constructing an adaptive dual-path network that bridges source and target domains. CyCADA \cite{hoffman2018cycada} combines spatial alignment of generated images with potential representation alignment to achieve unsupervised domain adaptation. In a different approach, DDMRL \cite{kim2019diversify} addresses object detection by employing two stages domain diversity and multi-domain invariant representation learning. WDASS \cite{das2023weakly} proposes a weakly-supervised domain adaptive semantic segmentation method that leverages prototypical contrastive learning to align features across domains.
Drawing inspiration from natural scene methods \cite{liu2021source,zhao2022source,xu2023multi}, domain adaptation has also been applied to medical image analysis \cite{chen2019synergistic}. DARCNN \cite{hsu2021darcnn} adapts models from natural scene images to hematoxylin-eosin (H\&E) stained images, capturing both domain-invariant and domain-specific features. CyC-PDAM \cite{liu2020unsupervised} transfers models between fluorescence images and H\&E images, while CAPL-Net \cite{li2022domain} focuses on domain adaptation for H\&E images from different sources. SAMDA \cite{wang2024samda} integrates the Segment Anything Model (SAM) with nnUNet \cite{isensee2021nnu} in the embedding space to enable few-shot domain adaptation. Despite the progress made by these domain adaptation methods, their performance often suffers due to insufficient understanding of target domain characteristics. In this paper, we propose leveraging auxiliary tasks to bridge the gap between the source domain model and the detection task, enhancing the efficiency of model transfer to the target domain.

\begin{figure*}
	\centering
	\includegraphics[width=7in]{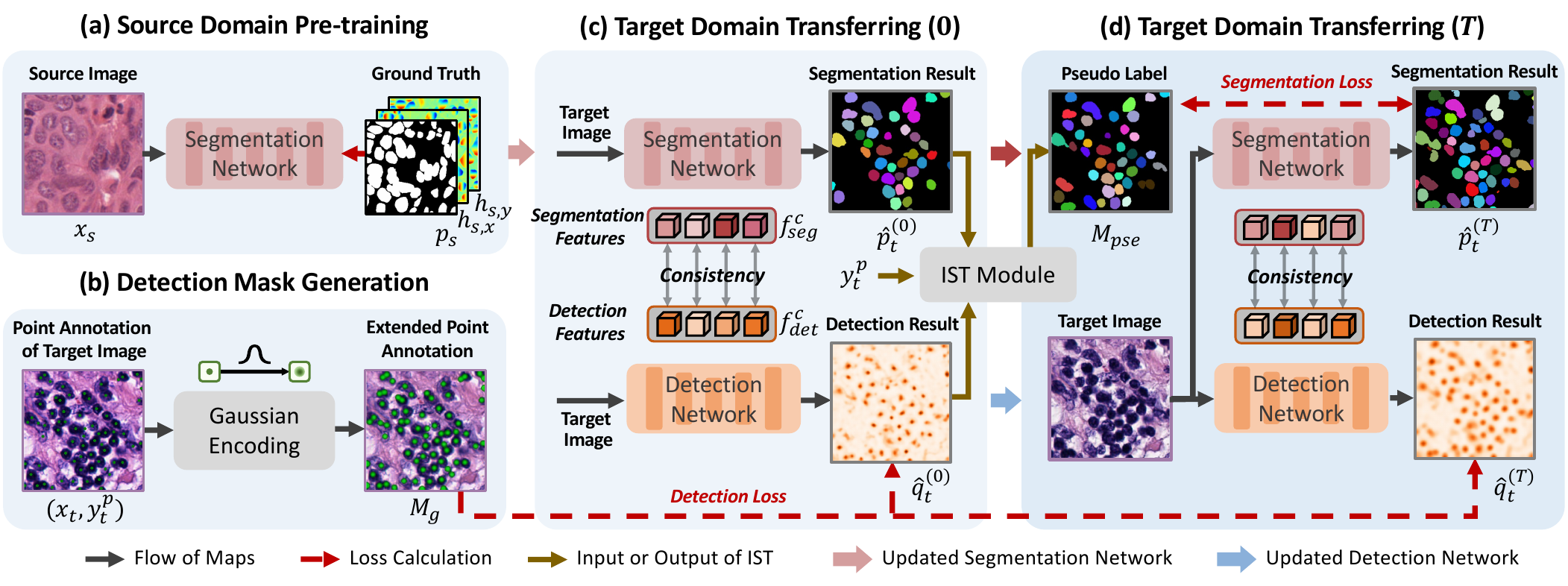}
	\caption{\textbf{The schematics of our proposed DAWN framework.} (a) represents the pre-trained segmentation network in source domain; (b) expresses the generation of extended point annotation using Gaussian distribution; (c) and (d) represent target domain transfer process. }
	\label{fig:flow chart}
\end{figure*}

\section{Methods}
\subsection{Framework Overview}\label{framework}

Detection and segmentation are closely related tasks that share semantic information, such as object location and texture, which can be leveraged to enhance both object detection and instance segmentation. In multi-task learning, detection and segmentation networks are often combined to improve the performance of both tasks \cite{he2017mask}. In the context of domain adaptation, we design a detection task using point annotations to support the segmentation network's adaptation to the target domain. By improving the segmentation network's adaptability to target domain data, this approach enhances nuclei segmentation accuracy, particularly in weakly supervised scenarios.

This paper introduces cross-task interaction strategies at both the feature and task levels. The feature-level interaction is designed to retain prior segmentation information relevant to the target domain. In contrast, task-level interaction compensates for the limitations of segmentation output through a combination of pseudo-label optimization and interactive supervision techniques. The overall flow of our proposed DAWN is illustrated in Fig. \ref{fig:flow chart}.

We first pre-train the segmentation network in the source domain, as illustrated in Fig. \ref{fig:flow chart} (a). To address the domain shift caused by variations in the data distribution, we introduce a Consistent Feature Constraint (CFC) module and a Combined Pseudo-Label (CPL) optimization module, as shown in Fig. \ref{fig:flow chart} (c)-(d). Additionally, to enhance the efficiency and effectiveness of domain transfer, we employ an Interactive Supervision Training (IST) method. Based on these components, our overall training loss function is defined as follows:
\begin{equation}\label{t_loss}
	Loss = L_{\text{det}} + \alpha L_{\text{fea}} + \beta L_{\text{dyn}},
\end{equation}
where $L_{\text{det}}$ represents the detection network loss, $L_{\text{fea}}$ is the consistent feature constraint loss, and $L_{\text{dyn}}$ corresponds to the dynamic supervision loss. The parameters $\alpha$ and $\beta$ are used to balance the weight of different loss terms. It should be noted that $L_{\text{det}}$ is associated with a fully supervised learning task.

\subsection{Consistent Feature Constraint}\label{cfc}
The semantic features encoded by detection and segmentation networks are correlated and complementary, which provides a basis for adaptive learning of the segmentation network with the help of an auxiliary detection network. The source domain dataset is defined as $D_{S}=\{(x_s,y_{s}^{f})\}_{s=1}^{m}$ and the target domain dataset is defined as $D_{T}=\{ (x_t,y_{t}^{p})\}_{t=1}^{n}$. $x_s$ and $x_t$ represent the input image of the source and target domain, respectively. $y_{s}^{f}$ represents fine pixel-level label and $y_{t}^{p}$ represents point label. $m$ and $n$ represent the number of images in the source and target domain, respectively. 

On source domain data $D_{S}$, we firstly pre-train a segmentation network as shown in Fig. \ref{fig:flow chart} (a) and Stage 1 of Algorithm \ref{algorithm1}. We adopt HoverNet \cite{graham2019hover} as a segmentation network in this step, and its supervision information consists of three components: foreground probability $p_s$, horizontal distance $h_{s,x}$, and vertical distance $h_{s,y}$. Using the pixel-level annotation $y_{s}^{f}$, the supervision images can be obtained through pre-processing, as shown in Fig. \ref{fig:flow chart} (a). In this step, the loss function \cite{graham2019hover} is formulated as follows:
\begin{equation}
	\begin{split}
		L_{pre} & = \text{CE}(\hat{p}_s, p_s)+\text{Dice}(\hat{p}_s, p_s) \\&
		+\text{MSE}(\hat{h}_{s,x}, h_{s,x})+\text{MSE}(\hat{h}_{s,y}, h_{s,y}) \\&
		+\text{MSE}(\nabla\hat{h}_{s,x}, \nabla h_{s,x})+\text{MSE}(\nabla \hat{h}_{s,y}, \nabla h_{s,y}) ,
	\end{split}\label{s_loss}
\end{equation}
where $\hat{p}_s$, $\hat{h}_{s,x}$, and $\hat{h}_{s,y}$ represent the predicted foreground probability, horizontal distance, and vertical distance. CE, Dice, and MSE represent cross-entropy loss, dice loss, and mean square error loss.

During the domain adaptation stage, we jointly train the segmentation and detection networks, and the process is shown in Fig. \ref{fig:flow chart} (c) and Stage 2 of Algorithm \ref{algorithm1}. First, the well-trained segmentation network is utilized to initialize the target domain segmentation model. And the encoder feature of the segmentation network is denoted as $f_{seg}$. Then, we map the feature $f_{seg}$ into a new feature space by a convolution layer and a fully connected layer and the output feature is denoted as $f_{seg}^{c}$ as follows:
\begin{equation}
    f_{seg}^{c} = FC(Conv(f_{seg})),
\end{equation}
where $FC$ and $Conv$ represent fully connected and convolution layers.
Finally, the output feature $f_{seg}^{c}$ is used to compute the consistent feature constraint.

In the detection network, considering single central pixel just represents nuclei position while ignoring nuclei texture, it is hard to train an effective detection network using target label $y_{t}^{p}$ \cite{paul2017count}. Thus, we employ an extended Gaussian distribution \cite{qu2020weakly} to model the morphology of nuclei. The extended Gaussian encoding process is shown in Fig. \ref{fig:flow chart} (b) and formulated as follows:
\begin{equation}\label{det_t}
	\begin{split}
		M_{g,i}= \left \{
		\begin{array}{ll}
			exp(-\frac{{D_i}^2}{2{\sigma}^2}), & \; D_i \leq r_1,\\
			0,     &  \; r_1< D_i \leq r_2, \\
			-1,    & otherwise,
		\end{array}
		\right.
	\end{split}
\end{equation}
where $D_i$ represents the distance between pixel $i$ and nearest point annotation, $r_1$ represents initialized nuclei radius, the region between $r_1$ and $r_2$ is background, $\sigma$ is the discrete degree of Gaussian distribution. In $M_{g,i}$, the pixel corresponding to $-1$ does not participate in the training. 
Employing the generated encoding, a supervised detection model is trained as follows:
\begin{equation}\label{det_loss}
	L_{det} = \frac{1}{|\Omega|} \sum_{i \in \Omega} w_i (\hat{q}_{t,i}^{(0)}-M_{g,i})^2,
\end{equation}
where $\Omega$ represents the pixel set participating training, \emph{i.e.} $\Omega=\{i|0<D_i\leq r_2\}$; $w_i$ is the training weight of pixel $i$. In order to pay more attention to the foreground, we set $w_i$ equal to $10$ when $D_i < r_1$; otherwise, $w_i$ equals $1$. 
The extended Gaussian encoding not only encodes tissue texture, but can also learn nuclei shape gradually, assisting in segmentation network optimization.

Similar to the above segmentation network, the encoder output $f_{det}$ of the detection network is mapped into new representation $f_{det}^{c}$, which will be used to calculate CFC with $f_{seg}^{c}$ as shown in Fig. \ref{fig:flow chart} (c) and the feature constraint loss $L_{fea}$ is calculated as follows:
\begin{equation}\label{fea_loss}
	L_{fea}=\frac{1}{|dim|} {\Vert f_{seg}^{c}-f_{det}^{c}\Vert}^2_2,
\end{equation}
where $dim$ represents the dimension of feature encoding. $f_{seg}^{c}$ and $f_{det}^{c}$ are feature embeddings of segmentation and detection networks. Due to the segmentation model being trained on the source domain dataset, its feature representation is closely related to the source domain. In contrast, the detection model is trained on the target domain, so its features are more relevant to the target domain data. Leveraging the target domain representation of $f_{det}^c$, the CFC effectively aligns $f_{seg}^c$ with $f_{det}^c$, enhancing the domain adaptation process.

\begin{algorithm}[t]
	\caption{The Whole Training Process of Our DAWN}
	\label{algorithm1}
	\begin{algorithmic}[1] 
		\STATE \textbf{Stage 1. Pre-train segmentation network}
		\STATE \textbf{Input:} Source domain data $D_s$;
		number of epoch $e_s$;
		\STATE \textbf{Output:} Network weight $W_s$;
		\FOR{epoch = 1,\dots, $e_s$}
		\STATE Calculate loss $L_{pre}$ in Eq. \ref{s_loss} and update network;
		\STATE Output: $\hat{p}_s$, $\hat{h}_{s,x}$, $\hat{h}_{s,y}$; 
		\ENDFOR
            \STATE Save weights: $W_s$.
		\STATE \textbf{Stage 2. Joint training networks}
		\STATE \textbf{Input:} Target domain data $D_t$; Point annotation $y_t^{p}$; initialized weight $W_{s}$;
		\STATE \textbf{Initialize:} $I=4$; $e_t=20$; Initialize segmentation network with $W_s$;
            \STATE \textbf{Output:} Instance segmentation results $\hat{y}_{t}$;
            \FOR{iter=1, \dots, $I$}
		\FOR{epoch=1,\dots,$e_t$}
		\STATE Calculate loss: $L_\text{det} + \alpha L_\text{fea}+\beta L_\text{dyn}$ in Eq. \ref{t_loss} and update network;
		\ENDFOR
            \STATE Prediction outputs: $\hat{p}_t$, $\hat{q}_t$; 
            \STATE Update pseudo-label: $M_{pse}$ $\leftarrow$ CPL($\hat{p}_t$, $ \hat{q}_t$, $y_t$);
		\STATE Training networks with $M_{pse}$ in Eq. \ref{loss seg};
            \STATE Post-process: $\hat{y}_{t}$ $\leftarrow$ Post($\hat{p}_t$, $\hat{h}_{t,x}$, $\hat{h}_{t,y})$;
            \ENDFOR
	\end{algorithmic}
\end{algorithm}

\subsection{Combined Pseudo-label Optimization}\label{dpl}

Due to domain shifts caused by variations in nuclei morphology and size across different organs, under-segmentation can occur in the target domain predictions. To address this issue, we propose a Combined Pseudo-Label (CPL) optimization method, which integrates the outputs of both the segmentation and detection networks. This approach utilizes the superior object location capabilities of the detection network to effectively compensate for the shortcomings of the segmentation network. A schematic diagram of the process is provided in Fig. \ref{fig:its}.

The CPL method comprises two key components: mask fusion and the distance filter. In mask fusion, the prediction outputs from both networks are combined to achieve comprehensive nuclei segmentation. The distance filter is then applied to refine the fused results by eliminating incorrectly segmented nuclei. The resulting outputs from the CPL serve as ground-truth annotations for supervising the subsequent training.

\subsubsection{Mask Fusion}
\begin{figure}[t]
	\centering
	\includegraphics[width=3.2in]{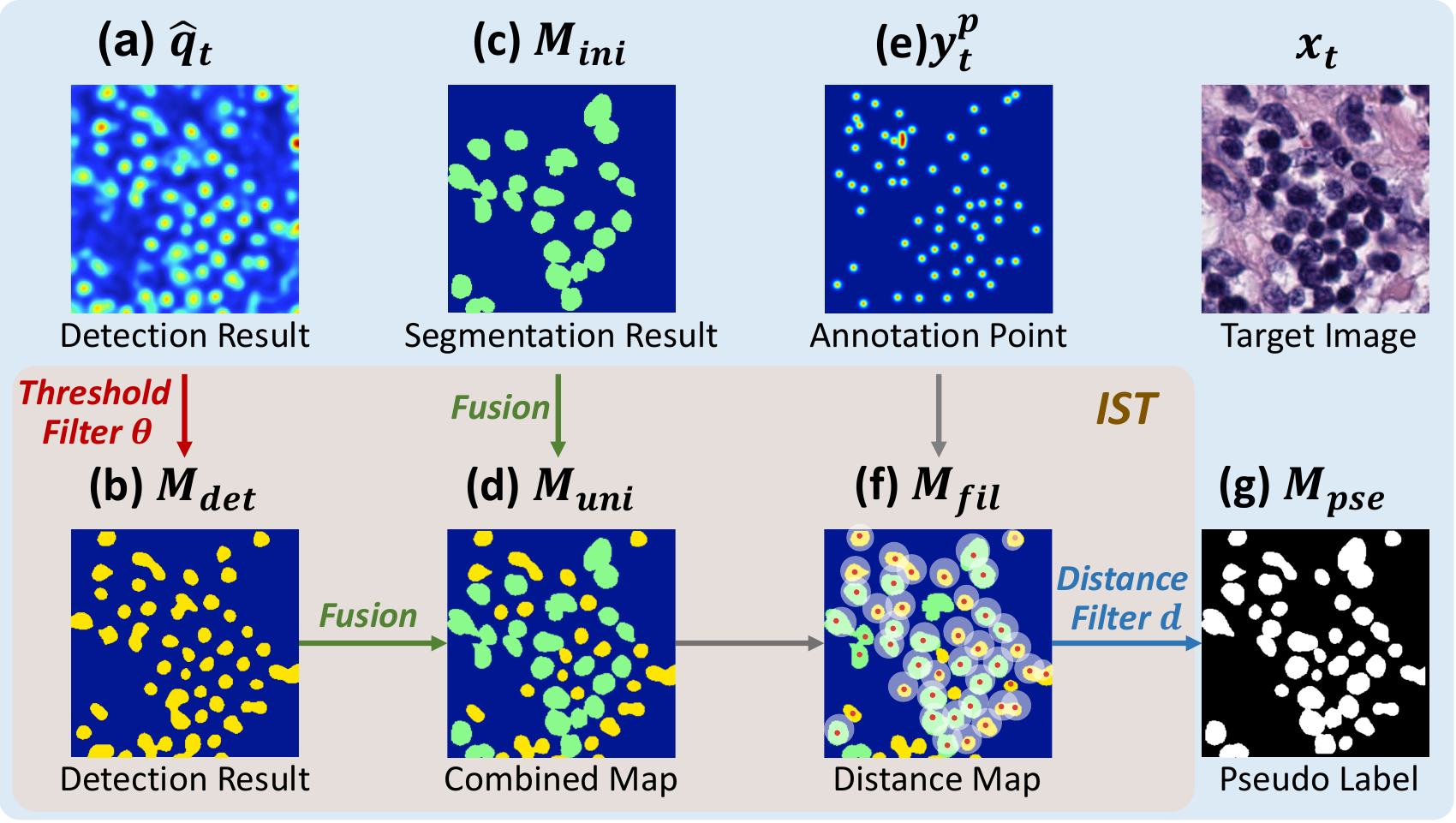}
	\caption{\textbf{The schematic diagram of combined pseudo-label optimization.} (a) is probability output of detection network; (b) is detection mask through filtering; (c) is segmentation mask; (d) is combined map of detection and segmentation outputs; (e) is point annotation; (f) is generated pseudo-label.}
	\label{fig:its}
\end{figure}

Based on the prediction output $\hat{q}_t$ of the detection network, we employ a threshold filter to preserve the high confidence region of prediction $\hat{q}_t$ as shown in Fig. \ref{fig:its} (a) and obtain the initialized detection $M_{det}$ as shown in Fig. \ref{fig:its} (b), which represents the nuclei region:
\begin{equation}\label{det_thre}
	\begin{aligned}
		& M_{det} = I_{[\hat{q}_t>\theta]}\in \{0,1\}^{N \times N},\\ 
	\end{aligned}
\end{equation}
where $\hat{q}_t$ represents the probability output of detection model, $\theta$ is the threshold value, $M_{det}$ represents generated nuclei region.
At the same time, we transform the probability output $\hat{p}_t$ of the segmentation network, obtaining the initialized $M_{ini}$, which has the under-segmentation problem, as shown in Fig. \ref{fig:its} (c). Then, we calculate a union set for $M_{ini}$ and $M_{det}$ as shown in Fig. \ref{fig:its} (d):
\begin{equation}
	M_{uni}=M_{ini} \cup M_{det}.
\end{equation}

In mask fusion, by utilizing the strong identification ability of $M_{det}$, we solve the under-segmentation problem led by the domain shift. The generated $M_{uni}$ contains the prediction result of $M_{ini}$ and a new mask in $M_{det}$.
It is worth noting that although detection output $M_{det}$ has an irregular nuclei shape, it will be optimized in the interactive supervision process, which will be stated in the following subsection.

\subsubsection{Distance Filter} We propose a distance filter method to solve incorrect segmentation in $M_{uni}$. This method utilizes point labels to locate the preserved nuclei and employs a distance-based metric to eliminate incorrectly segmented nuclei. Specifically, our method is described as follows. 

Firstly, we use the point annotation $y_{t}^{p}$ of the target domain data to locate the nuclei in the fused image $M_{uni}$. As illustrated in Fig. \ref{fig:its} (f), the red dots represent the annotated points. Next, we calculate the distance from each pixel to its nearest annotation point and choose pixels within a certain distance. In this process, the distance is set as $d$, and the pixels that have a distance smaller than $d$ are considered foreground, while other regions are treated as background. After the distance filter, we obtain the updated pseudo-label $M_{pse}$ as shown in Fig. \ref{fig:its} (g). The pseudo label $M_{pse}$ is a binary map; in the map, ``1" represents the foreground, and ``0" represents the background, which is used to guide the interactive training.

\subsection{Interactive Supervision Training Method}


This subsection proposes a dynamic pseudo-label optimization method to facilitate domain adaptation in the segmentation network. Specifically, the pseudo-labels generated from the CPL module in the $(t-1)^{th}$ round serve as supervision information for the training of segmentation and detection networks in the $t^{th}$ round. By iterative optimizing the segmentation network, this method enables progressive domain transfer. The interactive supervision process is shown in Fig. \ref{fig:flow chart} (c)-(d) and Algorithm \ref{algorithm1} Line 13-16.

In the $(t-1)^{th}$ round, the CPL module outputs the pseudo-labels, denoted as $M_{pse}$. The pseudo-label $M_{pse}$ combines the previous outputs of the segmentation and detection networks. Therefore, $M_{pse}$ is viewed as the supervision information for training the new round of the segmentation network, effectively addressing the under-segmentation issue in the segmentation results. At the same time, point annotation-based methods are used to supervise the detection network, which does not fully consider the morphological information of nuclei. Hence, we employ pseudo-labels to supervise the training of the detection network, which can accelerate the detection network to learn the nuclear morphology. During the training process, $M_{pse}$ always contains the prediction information from both the segmentation and detection networks. Thus, this method can be considered as an interactive supervision process, and the loss function for this method is shown below:
\begin{equation}\label{loss seg}
    L_{dyn} = \text{CE}(M_{pse}, \hat{p}_t) + \text{MSE}(M_{pse}, \hat{q}_t),
\end{equation}
where $M_{pse}$ is the pseudo-label generated by the CPL module, $\hat{p}_t$ and $\hat{q}_t$ represent the predicted output of segmentation and detection networks in the target domain.

By using the pseudo-labels to supervise the network training in a dynamic and progressive manner, our approach promotes the effectiveness of domain adaptation and enhances the performance of the segmentation network.

\begin{table}[htp]
	\centering
	\renewcommand{\arraystretch}{1.4}
	\caption{Hyper-parameter settings.}
	\begin{tabular}{cccccc}
		\hline
		\textbf{Dataset} & \textbf{$r_1$} & $r_2$ & $\sigma$  & $\theta$ & d  \\
		\hline
		\textbf{TNBC} & 11 & 22 & 2.75  & 0.2  & 25  \\
            \textbf{CryoNuSeg} & 11 & 22 & 2.75 & 0.2 & 25 \\
		\textbf{Lizard}  & 8 & 16 & 2  & 0.2 & 25  \\
		\textbf{ConSeP}  & 11 & 22 & 5 &  0.5 & 25 \\
		\hline
	\end{tabular}
	\label{tab:paras}
\end{table}

\section{Experiments}\label{exper}
\subsection{Datasets}
In this subsection, we introduce the PanNuke, MoNuSeg datasets used in the source domain (\ding{168}), and TNBC, CryoNuSeg, ConSeP datasets used in the target domain (\ding{171}). In detail, the PanNuke, MoNuSeg, TNBC, CryoNuSeg, and ConSeP datasets are sampled at ×40 magnification. The Lizard dataset is sampled at ×20 magnification.

\noindent \ding{168} \textit{\textbf{PanNuke}}\cite{gamper2020pannuke} comes from TCGA (The Cancer Genome Atlas), which consists of 205,343 labeled nuclei from 19 different organs in more than 2,000 visual fields. 

\noindent \ding{168} \textit{\textbf{MoNuSeg}}\cite{kumar2017dataset} comes from TCGA, which consists of 30 histopathology images from 7 different organs.

\noindent \ding{171} \textit{\textbf{TNBC}}\cite{naylor2018segmentation} comes from Curie Institute, which consists of 50 histopathology images from 11 breast cancer patients. 

\noindent \ding{171} \textit{\textbf{Lizard}}\cite{graham2021lizard} comes from UHCW (University Hospital Coventry and Warwickshire) and TCGA, which consists of 291 image regions from colorectal cancer patients.

\noindent \ding{171} \textit{\textbf{CryoNuSeg}}\cite{mahbod2021cryonuseg} comes from TCGA. The dataset consists of 30 histopathology images from 10 organs.

\noindent \ding{171} \textit{\textbf{ConSeP}}\cite{graham2019hover} comes from UHCW, which consists of 41 histopathology images from colon cancer patients.

\subsection{Implementation Details and Evaluation Metrics}
Our DAWN uses a U-Net framework with a ResNet \cite{he2016deep} backbone as a detection network. Meantime, a ResNet encoder combined with a DenseNet decoder is used as a segmentation network. All experiments are run with PyTorch on an Nvidia RTX 3090 GPU.

We set the loss weights in Eq. \ref{t_loss} to $\alpha=0.1$ and $\beta=0.15$. The hyperparameters specified in Eq. \ref{det_t} and Eq. \ref{det_thre} are detailed in Table \ref{tab:paras}. Specifically, $r_1$ and $\sigma$ are determined based on the mean and standard deviation of the nuclear radius in the validation set, with $r_2$ set at twice the value of $r_1$ to balance the foreground and background pixels. The parameter $\theta$ is established according to the output of the detection network in the validation set, while $d$ is determined empirically. We evaluate our experiments using the metrics of DICE, AJI\cite{kumar2017dataset}, DQ\cite{kirillov2019panoptic}, SQ\cite{kirillov2019panoptic} and PQ\cite{kirillov2019panoptic}. 

\begin{table*}[t]
\centering
\caption{The comparison of different weakly supervised and fully supervised methods.}
\renewcommand{\arraystretch}{1.3}
	\resizebox{\linewidth}{!}{
    \begin{tabular}{c|lccccc|c|lccccc}
        \Xhline{1.0pt}
        Datasets & Methods & DICE & AJI & DQ & SQ & PQ & Datasets & Methods & DICE & AJI & DQ & SQ & PQ \\
        \Xhline{1.0pt}
        & C2FNet\cite{tian2020weakly} & 0.754 & 0.605 &  0.763 & 0.682 & 0.505 & & C2FNet\cite{tian2020weakly} & 0.597 & 0.213 & 0.286  & 0.600 & 0.163  \\
        & DDTNet\cite{zhang2022ddtnet}	& 0.789	& 0.665	& \underline{0.807} & 0.745 & 0.601 & & DDTNet\cite{zhang2022ddtnet} & 0.602 & 0.236 & 0.350 & \underline{0.619} & 0.259 \\
        & BoNuS\cite{lin2024bonus} & \underline{0.833} & \textbf{0.712} & 0.798 & 0.743 & \textbf{0.635} & & BoNuS\cite{lin2024bonus} & 0.623 & 0.264 & \underline{0.364} & 0.617 & \underline{0.309} \\
        & WeakSeg\cite{qu2020weakly} & 0.811 & 0.647 & 0.804 & \underline{0.764} & 0.616 & & WeakSeg\cite{qu2020weakly} & \textbf{0.714} & \textbf{0.381} & \textbf{0.403} & \textbf{0.676} & \textbf{0.334} \\
        & MaskGANet\cite{guo2021learning} & 0.751 & 0.597 & 0.744	& 0.612	& 0.454 & & MaskGANet\cite{guo2021learning} & 0.603 & 0.197 & 0.256 & 0.584 & 0.134 \\
        TNBC & SPF-TN \cite{zhang2018spftn} & 0.806 & 0.687 & \textbf{0.812} & 0.739 & 0.611 & Lizard & SPF-TN\cite{zhang2018spftn} & 0.610 & 0.221 & 0.314 & 0.602 & 0.261 \\
         (a)& PENet \cite{yoo2019pseudoedgenet} & 0.682 & 0.497 & 0.688 & 0.547 & 0.394 & (c) & PENet \cite{yoo2019pseudoedgenet} & 0.545 & 0.140 & 0.229 &  0.534 & 0.105 \\
         & WeakSAM\cite{zhu2024weaksam} & \textbf{0.834} & \underline{0.697} & 0.795 & \textbf{0.782} & \underline{0.620} &  & WeakSAM\cite{zhu2024weaksam} & \underline{0.643} & \underline{0.279} & 0.314 & 0.608 & 0.267  \\
        \cline{2-7}  \cline{9-14}
        & U-Net \cite{ronneberger2015u}	& 0.783 & 0.692	& 0.754 & 0.750	& 0.617 & & U-Net \cite{ronneberger2015u}	& 0.716 & 0.365 & 0.436 & 0.633 & 0.353 \\
        & HoverNet \cite{graham2019hover} & 0.815 & 0.662 & 0.793 & 0.811	& 0.644 & & HoverNet\cite{graham2019hover} & \textbf{0.803} & 0.465 & 0.573	& \textbf{0.780} & \underline{0.453}	\\
        & CDNet \cite{he2021cdnet}	& \textbf{0.847} & \textbf{0.724} & \textbf{0.840} & \textbf{0.819} & \textbf{0.684} & & CDNet \cite{he2021cdnet} & 0.708 & 0.507 & \textbf{0.630} & \underline{0.764} & \textbf{0.487} \\
        & DCAN \cite{chen2016dcan}	& 0.704	&0.497	&0.623	&0.661	&0.433	& & DCAN \cite{chen2016dcan} & 0.716 & 0.290 & 0.435 & 0.632 & 0.277 \\
        & Dist \cite{naylor2018segmentation} & \underline{0.836} & 0.681 & 0.770 & 0.792 & 0.611 & & Dist \cite{naylor2018segmentation} & 0.753 & \textbf{0.520} & 0.588	& 0.676	& 0.423 \\
        & Micro-Net \cite{raza2019micro} & 0.813 & 0.673
        & 0.772	& 0.795	& 0.615 & & Micro-Net \cite{raza2019micro} & \underline{0.776} & \underline{0.519} & \underline{0.615} & 0.759	& 0.451 \\
        & AOST \cite{yang2024scalable} & 0.821 & \underline{0.694} & \underline{0.822} & \underline{0.814} & \underline{0.674} & & AOST\cite{yang2024scalable} & 0.709 & 0.394 & 0.523	& 0.712	& 0.364 \\
        \cline{2-7}  \cline{9-14}
        & \textbf{DAWN (Ours)}	&\textbf{0.856}	&\textbf{0.728}	&\textbf{0.843}	&\textbf{0.820}	&\textbf{0.693}	& &  \textbf{DAWN (Ours)} & \textbf{0.728} & \textbf{0.403}	&\textbf{0.510}	&\textbf{0.736}	&\textbf{0.377} \\
        \Xhline{1.0pt}
        Datasets & Methods & DICE & AJI & DQ & SQ & PQ & Datasets & Methods & DICE & AJI & DQ & SQ & PQ \\
        \Xhline{1.0pt}
        & C2FNet\cite{tian2020weakly} &  0.643 & 0.384 & 0.581 & 0.573 & 0.332 & &  C2FNet\cite{tian2020weakly} & \textbf{0.749} & \textbf{0.464} & \textbf{0.497} & \textbf{0.706} & \underline{0.398}  \\
        & DDTNet\cite{zhang2022ddtnet} & \underline{0.696} & 0.440 & \underline{0.618}	& \underline{0.668} & 0.389 & & DDTNet\cite{zhang2022ddtnet} & 0.598	& 0.295	& 0.282	& 0.452	& 0.186 \\
        & BoNuS\cite{lin2024bonus} & 0.693 & 0.431 & 0.607 & 0.645 & \underline{0.399} & & BoNuS\cite{lin2024bonus} & 0.651 & 0.354 & 0.378 & 0.675 & 0.380  \\
        & WeakSeg\cite{qu2020weakly} & 0.682 & 0.410 & 0.579 & 0.649 & 0.357 &  & WeakSeg\cite{qu2020weakly} & 0.646 & 0.366 & 0.363 & 0.669 & 0.391 \\
        & MaskGANet\cite{guo2021learning} & 0.647 & 0.393 & 0.595 & 0.654 & 0.380 & & MaskGANet\cite{guo2021learning} & 0.653 & 0.367 & \underline{0.401} & \underline{0.696} & 0.374 \\
        CryoNuSeg & SPF-TN \cite{zhang2018spftn} & 0.676 & \underline{0.449} & 0.611 & 0.634 & 0.391 & ConSeP & SPF-TN\cite{zhang2018spftn} & 0.632 & \underline{0.392} & 0.373 & 0.667 & 0.382 \\
         (b) & PENet\cite{yoo2019pseudoedgenet} & 0.620 & 0.321 & 0.507 & 0.529 & 0.306 & (d) & PENet \cite{yoo2019pseudoedgenet} & 0.569 & 0.226 & 0.245 & 0.307 & 0.171 \\
          & WeakSAM\cite{zhu2024weaksam} & \textbf{0.754} & \textbf{0.480} & \textbf{0.627} & \textbf{0.685} & \textbf{0.449} & & WeakSAM\cite{zhu2024weaksam} & \underline{0.669} & 0.378 & 0.384 & 0.680 & \textbf{0.401} \\
         \cline{2-7}  \cline{9-14}
         & U-Net \cite{ronneberger2015u}	& 0.697 & 0.469	& 0.546 & \underline{0.731} & 0.403 & & U-Net \cite{ronneberger2015u} & 0.761 & 0.499 & 0.569 & 0.725 & 0.434 \\
         & HoverNet \cite{graham2019hover} & \textbf{0.804} & \underline{0.526} & \underline{0.655} & \textbf{0.754} & \underline{0.495} & & HoverNet \cite{graham2019hover} & \textbf{0.837} & 0.513 & \underline{0.636} & \underline{0.767} & \underline{0.492} \\
        & CDNet \cite{he2021cdnet}	& 0.776 & \textbf{0.539} & \textbf{0.660} & \textbf{0.754} & \textbf{0.499} & & CDNet \cite{he2021cdnet} & \underline{0.835} & \textbf{0.541} & \textbf{0.652} & \textbf{0.771} & \textbf{0.514} \\
        & DCAN \cite{chen2016dcan}	&0.766	&0.335	& 0.379	& 0.681	&0.258	& & DCAN \cite{chen2016dcan} & 0.745 & 0.312 & 0.439	& 0.683	& 0.300 \\ 
	& Micro-Net \cite{raza2019micro} & 0.690 & 0.435
        & 0.463	& 0.701	& 0.327 & & Micro-Net \cite{raza2019micro} & 0.805 & \underline{0.516} & 0.602 & 0.755 & 0.457 \\
        & Dist \cite{naylor2018segmentation} & \underline{0.792} & 0.518	&0.595	&0.721	&0.432	& & Dist\cite{naylor2018segmentation} & 0.793 & 0.512	&0.587	&0.716	& 0.422 \\
        & AOST \cite{yang2024scalable} & 0.754 & 0.487 & 0.602 & 0.724 & 0.459 & & AOST\cite{yang2024scalable} & 0.764 & 0.478 & 0.565 & 0.704 & 0.420\\
        \cline{2-7}  \cline{9-14}
        \cline{2-7}  \cline{9-14}
        & \textbf{DAWN (Ours)}	&\textbf{0.804}	& \textbf{0.508} &\textbf{0.637}	&\textbf{0.744}	&\textbf{0.476}
        & & \textbf{DAWN (Ours)} &\textbf{0.805} &\textbf{0.509}	&\textbf{0.627}	&\textbf{0.759}	&\textbf{0.477}\\
        \Xhline{1.0pt}
    \end{tabular}}
    \label{tab:compare}
\end{table*}

\subsection{Comparison with Weakly Supervised Methods}

We compare our proposed domain-adaptive weakly supervised nuclei segmentation method with eight other models on four datasets. The comparison methods consist of DDTNet\cite{zhang2022ddtnet}, BoNuS\cite{lin2024bonus}, WeakSeg\cite{qu2020weakly},  MaskGANet\cite{guo2021learning}, C2FNet\cite{tian2020weakly}, PENet \cite{yoo2019pseudoedgenet}, SPF-TN \cite{zhang2018spftn}, and WeakSAM\cite{zhu2024weaksam}. The experiment results are shown in Table \ref{tab:compare}. In the table, the \textbf{best} weakly supervised method is bold, and the \underline{second best} weakly supervised method is underlined. 
For a fair comparison, we use the same pseudo labels as our DAWN for the two-stage methods WeakSeg, BoNuS, and DDTNet. In detail, the pre-trained segmentation network is tested on the target dataset, and the prediction outputs are used as pseudo labels. In the experiment, we use PanNuke as the source domain dataset. 

\subsubsection{Comparisons on Breast Cancer TNBC}

We validate our proposed DAWN on a breast cancer dataset, and the experimental results are shown in Table \ref{tab:compare} (a). From this table, we observe that our proposed method outperforms existing state-of-the-art weakly supervised segmentation methods. Compared with the best method, our DAWN achieves improvements of 2.2\% and 1.6\% in terms of DICE and AJI metrics, respectively. Moreover, our method exhibits significant advantages over traditional two-stage weakly supervised models such as MaskGANet and WeakSeg. In detail, our method surpasses MaskGANet by at least 10\% and surpasses WeakSeg by at least 4.0\% on all five metrics.

Furthermore, we conduct visualization comparison experiments as shown in the first two rows of Fig. \ref{fig:weak_com}. From the segmentation results, it can be seen that our method has clear advantages. On the TNBC and CryoNuSeg datasets, our method segments cell nuclei contours more accurately, while on the Lizard and ConSeP datasets, when the nuclei distribution is dense, our method can more comprehensively identify nuclei.

\subsubsection{Comparisons on Pan-Cancer CryoNuSeg}

We evaluate the performance of our proposed method, DAWN, on the challenging pan-cancer CryoNuSeg dataset. This dataset presents significant variations in nuclear morphology, posing substantial challenges for training. Our experimental results, as shown in Table \ref{tab:compare} (b), demonstrate that DAWN outperforms the leading method, WeakSAM, with improvements of 5.0\% and 2.8\% in DICE and AJI metrics, respectively. Additionally, DAWN shows notable gains over the second-best approach, achieving improvements of approximately 10\%, 6\%, and 8\% in DICE, AJI, and PQ metrics, respectively.

We perform visual comparison experiments, and the results are shown in the third and fourth rows of Fig. \ref{fig:weak_com}. The segmentation results of WeakSeg, MaskGANet, and C2FNet show that these methods exhibit poorer segmentation results when faced with significant morphological variations of the nuclei. In contrast, our method still achieves outstanding performance. We attribute this phenomenon to the fact that two-stage methods struggle to adapt their pseudo-labels to the variations in staining and morphology, leading to lower segmentation performance. However, our method utilizes a domain transfer strategy that leverages prior knowledge from the source domain and introduces a progressive domain adaptation method to adapt to the target domain's image characteristics.

\subsubsection{Comparisons on Colorectal Cancer Lizard and ConSeP}

\begin{figure*}[t]
	\centering
	\includegraphics[width=7.0in]{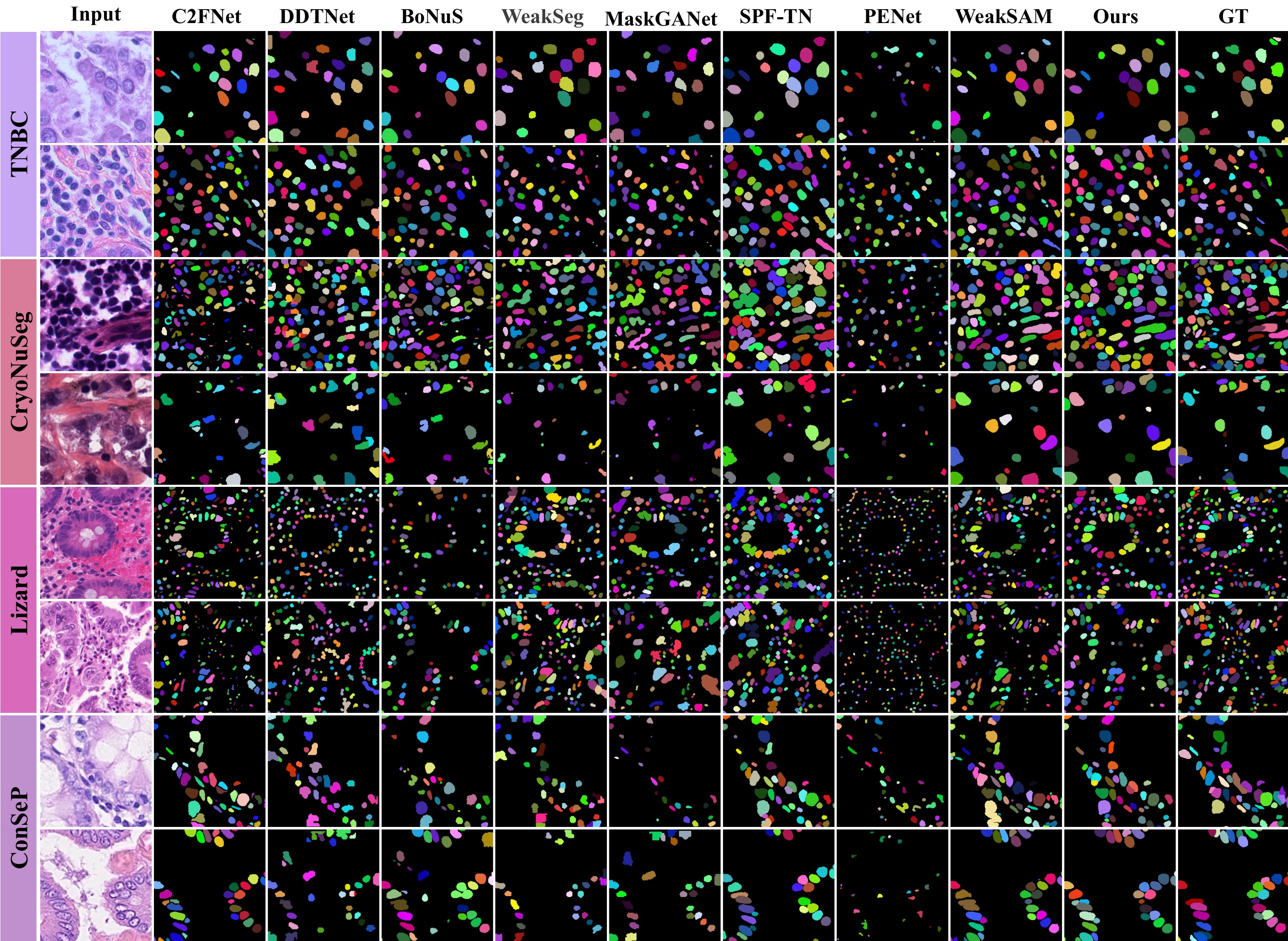}
	\caption{The visualization comparison between weakly supervised methods and our proposed DAWN.}
	\label{fig:weak_com}
\end{figure*}

We further validate the performance of our proposed method in pathological images of colorectal cancer. As shown by the results of the Lizard dataset in Table \ref{tab:compare} (c), existing weakly supervised methods, particularly C2FNet, MaskGANet, and PENet, perform poorly, with PQ scores of 0.134, 0.163, and 0.105, respectively. This underperformance can be attributed to the fact that the Lizard dataset is sampled at a magnification of $\times 20$, resulting in densely distributed and small nuclei, which presents significant challenges for segmentation. Despite these difficulties, our method achieves improvements over the best-performing method, WeakSeg, with increases of 1.4\% and 2.2\% in DICE and AJI metrics, respectively. Furthermore, as illustrated in the fifth and sixth rows of Fig. \ref{fig:weak_com}, our method maintains excellent segmentation performance even with smaller nuclei. In contrast, the segmentation results of DDTNet, which performed well on the TNBC and CryoNuSeg datasets, are unsatisfactory on Lizard. This highlights that our method is less affected by the sampling magnification compared to other models.

As shown in the ConSeP results in Table \ref{tab:compare} (d), our proposed DAWN model achieves significant improvements over other weakly supervised models. Specifically, DAWN outperforms C2FNet with gains of 5.6\% in DICE and 4.5\% in AJI. The visual results in the last two rows of Fig. \ref{fig:weak_com} highlight that the nuclei in the ConSeP dataset often exhibit blurry morphology and are densely distributed, making it difficult for other methods to accurately delineate boundaries and distinguish individual nuclei instances. In contrast, our method addresses these challenges more effectively, leading to improved segmentation performance.

\subsection{Comparison with Fully Supervised Methods}

To compare the performance difference between our proposed weakly supervised segmentation method and existing fully supervised segmentation methods, we compare our DAWN with seven state-of-the-art models, including U-Net \cite{ronneberger2015u}, HoverNet \cite{graham2019hover}, CDNet \cite{he2021cdnet}, DCAN \cite{chen2016dcan}, Micro-Net \cite{raza2019micro}, Dist \cite{naylor2018segmentation}, and AOST\cite{yang2024scalable}. In the experiment, we use PanNuke as the source domain and TNBC and ConSeP as the target domains. For a fair comparison, all comparison models are pre-trained in PanNuke and then fine-tuned in the four datasets. The results are shown in Table \ref{tab:compare}. In the table, the \textbf{best} fully supervised method is bold, and the \underline{second best} fully supervised method is underlined. 

\subsubsection{Quantitative Comparisons}

As shown in Table \ref{tab:compare}, our method achieves performance comparable to fully supervised methods on the TNBC dataset. Specifically, our method surpasses the best-performing CDNet by 0.9\% in DICE and significantly outperforms other traditional supervised methods such as U-Net and DCAN. On the CryoNuSeg and ConSeP datasets, although our method does not achieve the top performance, it is only 1.8\% and 0.4\% lower than the second-best HoverNet in terms of AJI. However, our method underperforms on the Lizard dataset. While it outperforms U-Net and DCAN, a performance gap remains compared to other methods. We attribute this to the inherent limitations of weakly supervised tasks, as using point annotations for domain transfer continues to present challenges.

\subsubsection{Qualitative Comparisons}
\begin{figure*}[ht]
	\centering
	\includegraphics[width=6.8in]{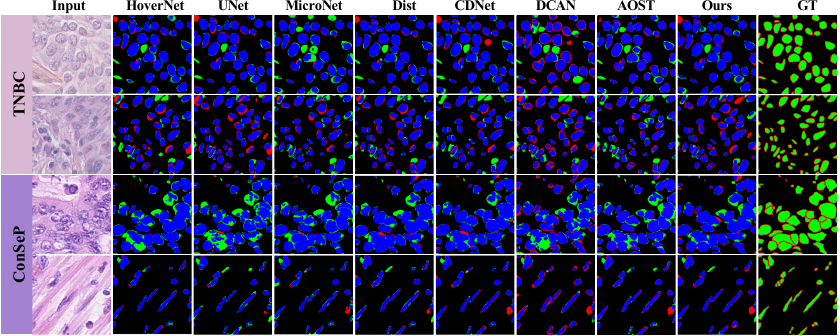}
	\caption{\textbf{The segmentation comparison between fully supervised methods and our proposed DAWN.} Blue stands for correctly true positive pixels, red stands for false positive pixels and green stands for false negative pixels.}
	\label{fig:comp1}
\end{figure*}

Furthermore, we provide comparison visualization in Fig. \ref{fig:comp1}. In the figure, blue stands for correctly true positive pixels, red stands for false positive pixels and green stands for false negative pixels. From the figure we can see our weakly supervised segmentation method achieves comparable performance with supervised methods. Surprisingly, our method even outperforms Micro-Net and U-Net. This indicates an accurate pseudo label initialization and an elaborate pseudo label optimization also provide fine supervision information although without pixel-level mask.

\begin{table}[t]
\centering
\caption{The domain adaptation comparison experiments from PanNuke dataset to TNBC and CryoNuSeg datasets.}
\renewcommand{\arraystretch}{1.15}
    \begin{tabular}{c|ccccc}
        \hline 
        \multirow{2}{*}{\textbf{Methods}} & \multicolumn{5}{c}{\textbf{PanNuke}$\rightarrow$\textbf{TNBC}} \\ 
        \cline{2-6}
        & \textbf{DICE} & \textbf{AJI} &\textbf{DQ} & \textbf{SQ} & \textbf{PQ} \\
        \hline
        \textbf{source only} & 0.8216 & 0.6764 & 0.8158 & 0.8094 & 0.6610 \\
        \textbf{SIFA} \cite{chen2019synergistic} & 0.7934 & 0.6520 & 0.7827 & 0.7493 & 0.6291 \\
        \textbf{PDAM} \cite{liu2020pdam} & 0.8192	& 0.6733 & 0.8104 & 0.7831 & 0.6577 \\
        \textbf{SFDA-CellSeg} \cite{li2023towards} & 0.8340 & 0.6817 & 0.8025 & 0.7933 & 0.6715 \\
        \textbf{ADPL}\cite{cheng2023adpl} & \underline{0.8409} & \underline{0.7122} & 0.8193 & \underline{0.8117} & \underline{0.6746} \\
        \textbf{CyCADA}\cite{hoffman2018cycada} & 0.8093 & 0.6712 & \underline{0.8249} & 0.7677 & 0.6343 \\
        \textbf{DDMRL}\cite{kim2019diversify} & 0.8141 & 0.6869 & 0.8225 & 0.7763 & 0.6509 \\
        \textbf{Ours} &\textbf{0.8561}	&\textbf{0.7280} & \textbf{0.8425} & \textbf{0.8202} & \textbf{0.6925}	 \\
        \hline
        \multirow{2}{*}{\textbf{Methods}} & \multicolumn{5}{c}{\textbf{PanNuke}$\rightarrow$\textbf{CryoNuSeg}} \\ 
        \cline{2-6}
        & \textbf{DICE} & \textbf{AJI} &\textbf{DQ} & \textbf{SQ} & \textbf{PQ} \\
        \hline
        \textbf{source only} & 0.7857 & 0.4699 & 0.6121 & \textbf{0.7467} & \underline{0.4667} \\
        \textbf{SIFA} \cite{chen2019synergistic} & 0.7437 & 0.4526 & 0.5731 & 0.7012 & 0.4011 \\
        \textbf{PDAM}\cite{liu2020pdam} & 0.7855 & 0.4748 & 0.6030 & 0.7381 & 0.4461 \\
        \textbf{SFDA-CellSeg} \cite{li2023towards} & 0.7912 & 0.4630 & 0.6121 & 0.7277 & 0.4480 \\
        \textbf{ADPL}\cite{cheng2023adpl} & 0.7847 & \underline{0.4801} & 0.6104 & 0.7333 & 0.4542 \\
        \textbf{CyCADA}\cite{hoffman2018cycada} & \underline{0.7993} & 0.4794 & \underline{0.6208} & 0.7278 & 0.4534 \\
        \textbf{DDMRL}\cite{kim2019diversify} & 0.7764 & 0.4692 & 0.5998 & 0.7194 & 0.4449 \\
        \textbf{Ours} &\textbf{0.8041}	& \textbf{0.5079} & \textbf{0.6374}	& \underline{0.7439}	&\textbf{0.4760} \\
        \hline
    \end{tabular}
    \label{tab:da1}
\end{table}

\subsection{Domain Adaptation Evaluation}
We conduct a domain adaptation comparison with six domain adaptation methods: SIFA\cite{chen2019synergistic}, PDAM\cite{liu2020pdam} and SFDA-CellSeg\cite{li2023towards}, ADPL\cite{cheng2023adpl}, CyCADA\cite{hoffman2018cycada}, and DDMRL\cite{kim2019diversify}. The domain adaptation experiments are designed from the following three perspectives.

\subsubsection{PanNuke as Source dataset}
To compare the different domain adaptation methods, we use a large pan-cancer dataset, PanNuke, as the source domain data and TNBC (breast cancer) and CryoNuSeg (pan-cancer) as the target domain data for the experiment. The experiment results are shown in Table \ref{tab:da1}. From this table, we can observe that our method achieves significant improvements both on TNBC and CryoNuSeg datasets. Regarding the AJI metric, our method outperforms the second-best method by 1.58\% and 2.78\% on TNBC and CryoNuSeg, respectively.

\begin{table}[t]
\centering
\caption{The domain adaptation comparison experiments from MoNuSeg dataset to TNBC and CryoNuSeg datasets.}
\renewcommand{\arraystretch}{1.15}
	\begin{tabular}{c|ccccc}
		\hline %
		\multirow{2}{*}{\textbf{Methods}} & \multicolumn{5}{c}{\textbf{MoNuSeg}$\rightarrow$\textbf{TNBC}} \\ 
		\cline{2-6}
		& \textbf{DICE} & \textbf{AJI} &\textbf{DQ} & \textbf{SQ} & \textbf{PQ} \\
		\hline
            \textbf{source only} & 0.7574 & 0.6323 & 0.7515 & 0.7333 & 0.6017 \\
		\textbf{SIFA} \cite{chen2019synergistic} & 0.7530 & 0.6289 & 0.7304 &  0.7157 & 0.5858 \\
		\textbf{PDAM} \cite{liu2020pdam}& 0.7873 & 0.6421 & 0.7511 & 0.7543 & 0.6117 \\
		\textbf{SFDA-CellSeg} \cite{li2023towards} & 0.8007 & 0.6597 & 0.7404 & 0.7484 & 0.6301 \\
            \textbf{ADPL}\cite{cheng2023adpl} & 0.8169 & \underline{0.6723} & \underline{0.7677} & \underline{0.7648} & \underline{0.6501} \\
        \textbf{CyCADA}\cite{hoffman2018cycada} & 0.8204 & 0.6619 & 0.7549 & 0.7627 & 0.6480  \\
        \textbf{DDMRL}\cite{kim2019diversify} & \underline{0.8211} & 0.6544 & 0.7577 & 0.7612 & 0.6419  \\
		\textbf{Ours} &\textbf{0.8294}	&\textbf{0.6933} & \textbf{0.8074} & \textbf{0.8085} & \textbf{0.6679}	 \\
		\hline
		\multirow{2}{*}{\textbf{Methods}} & \multicolumn{5}{c}{\textbf{MoNuSeg}$\rightarrow$\textbf{CryoNuSeg}} \\ 
		\cline{2-6}
		& \textbf{DICE} & \textbf{AJI} &\textbf{DQ} & \textbf{SQ} & \textbf{PQ} \\
		\hline
            \textbf{source only} & 0.7611 & 0.4534 & 0.5698 & \textbf{0.7221} & 0.4002 \\
		\textbf{SIFA} \cite{chen2019synergistic} & 0.7404 & 0.4421 & 0.5671 & 0.6924 & 0.3988 \\
		\textbf{PDAM} \cite{liu2020pdam} & 0.7679 & \underline{0.4654} & 0.5899 & 0.7121 & 0.4234 \\
		\textbf{SFDA-CellSeg} \cite{li2023towards} & \underline{0.7687} & 0.4521 & 0.6004 & 0.7133 & \underline{0.4287} \\
            \textbf{ADPL}\cite{cheng2023adpl} & 0.7527 & 0.4644 & 0.5998 & 0.7107 & 0.4250 \\
        \textbf{CyCADA}\cite{hoffman2018cycada} & 0.7622 & 0.4548 & \underline{0.6010} & 0.7109 & 0.4257 \\
        \textbf{DDMRL}\cite{kim2019diversify} & 0.7601 & 0.4547 & 0.5989 & 0.7034  & 0.4241 \\
		\textbf{Ours} &\textbf{0.7864}	& \textbf{0.4845} & \textbf{0.6122}	& \underline{0.7216}	&\textbf{0.4498} \\
		\hline
	\end{tabular}
	\label{tab:da2}
\end{table}

\subsubsection{MoNuSeg as Source dataset}
To validate the influences of data scale, we conducted domain adaptation experiments on a smaller pan-cancer dataset, MoNuSeg. The experiment results are shown in Table \ref{tab:da2}. When using MoNuSeg as the source domain data, the AJI metric of our method surpasses the second-best method by 2.10\% and 1.91\% on TNBC and CryoNuSeg target data, respectively. This indicates that our method is effective on small-scale datasets. Besides, comparing the results in Table \ref{tab:da1} and Table \ref{tab:da2}, the latter's performance significantly decreases. On the CryoNuSeg data, the DICE of the ADPL and CyCADA decreased by 3.20\% and 3.71\%, while our proposed method decreased by only 1.77\%, showing a smaller decline compared to the ADPL and CyCADA. The phenomenon suggests that our method is less affected by the data scale disturbance.

\begin{figure*}[t]
	\centering
	\includegraphics[width=6.6in]{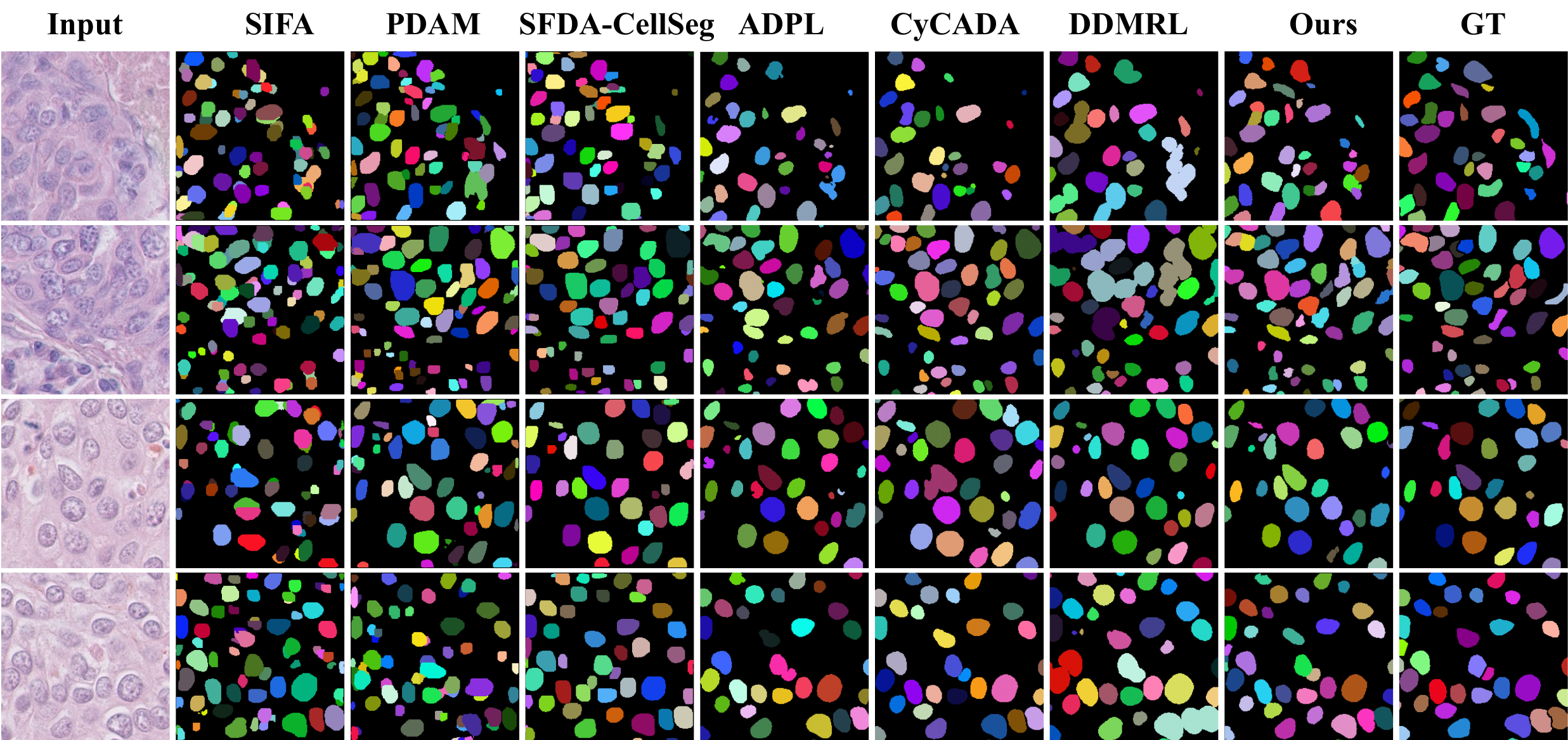}
	\caption{The visualization comparisons between three domain adaptation methods and our proposed DAWN from PanNuke dataset to TNBC dataset. }
	\label{fig:domain}
\end{figure*}

\begin{table}[t!]
\centering
\caption{The cross-cancer domain adaptation experiments from ConSeP (TNBC) dataset to TNBC (ConSeP) dataset.}
\renewcommand{\arraystretch}{1.15}
    \begin{tabular}{c|ccccc}
        \hline
        \multirow{2}{*}{\textbf{Methods}} & \multicolumn{5}{c}{\textbf{ConSeP}$\rightarrow$\textbf{TNBC}} \\ 
        \cline{2-6}
        & \textbf{DICE} & \textbf{AJI} &\textbf{DQ} & \textbf{SQ} & \textbf{PQ} \\
        \hline
        \textbf{source only} & 0.7544 & 0.6035 & \underline{0.7144} &  0.6961 & 0.5736  \\
        \textbf{SIFA} \cite{chen2019synergistic} & 0.7101 & 0.5414 & 0.5528 & 0.7012 & 0.5413 \\
        \textbf{PDAM} \cite{liu2020pdam} & 0.7302  & 0.5661 & 0.5757 & 0.7279 & 0.5725 \\
        \textbf{SFDA-CellSeg} \cite{li2023towards} & 0.7582 & 0.6081 & 0.6033 & 0.7322 & 0.5921 \\
        \textbf{ADPL}\cite{cheng2023adpl} & \underline{0.7696} & \underline{0.6231} & 0.6547 & \underline{0.7597} & \underline{0.6120} \\
        \textbf{CyCADA}\cite{hoffman2018cycada} & 0.7654 & 0.6209 & 0.6383 & 0.7404 & 0.6001  \\
        \textbf{DDMRL}\cite{kim2019diversify} & 0.7585 & 0.6199 & 0.6217 & 0.7383 & 0.5997 \\
        \textbf{Ours} &\textbf{0.8285} &\textbf{0.6634} & \textbf{0.7894} & \textbf{0.7855} & \textbf{0.6453}	 \\
        \hline
        \multirow{2}{*}{\textbf{Methods}} & \multicolumn{5}{c}{\textbf{TNBC}$\rightarrow$\textbf{ConSeP}} \\ 
        \cline{2-6}
        & \textbf{DICE} & \textbf{AJI} &\textbf{DQ} & \textbf{SQ} & \textbf{PQ} \\
        \hline
        \textbf{source only} & 0.6879 & 0.4391 & \underline{0.5273} & 0.6400 & 0.3924  \\
        \textbf{SIFA} \cite{chen2019synergistic}  & 0.6786  & 0.4076 & 0.4400 & 0.6592 & 0.3657 \\
        \textbf{PDAM} \cite{liu2020pdam} &  0.6829 & 0.4324 & 0.4590 & \underline{0.6782} & 0.3712 \\
        \textbf{SFDA-CellSeg} \cite{li2023towards} & 0.6913 & 0.4217 & 0.4484 & 0.6520 & 0.3817 \\
        \textbf{ADPL}\cite{cheng2023adpl} & 0.7174 & \underline{0.4437} & 0.4824 & 0.6634 & 0.3883 \\
        \textbf{CyCADA}\cite{hoffman2018cycada} & \underline{0.7274} & 0.4430  & 0.4746 & 0.6565 & \underline{0.3943}  \\
        \textbf{DDMRL}\cite{kim2019diversify} & 0.7107 & 0.4349 & 0.4680 & 0.6443  & 0.3814  \\
        \textbf{Ours} &\textbf{0.7743}	&\textbf{0.4607} &\textbf{0.5549} & \textbf{0.6981} & \textbf{0.4168}  \\
        \hline
    \end{tabular}
    \label{tab:da3}
\end{table}

\subsubsection{Cross-cancer Domain Adaptation}
We conduct cross-cancer domain adaptation experiments on the ConSeP (colon cancer) and TNBC (breast cancer) datasets, and the experiment results are shown in Table \ref{tab:da3}. We can observe that our method achieves the best performance among these methods. Particularly on the TNBC, the AJI surpasses the second-best method by 4.03\%. These experiments demonstrate the significant advantage of our method. Based on this phenomenon, we can conclude that our method is least affected by the source domain data. Our method can still exploit essential prior information for different cancer types to enhance the segmentation performance. 


\subsubsection{Qualitative Comparisons}
We provide visualization comparisons of segmentation results between domain adaptation methods in Fig. \ref{fig:domain}. First, our model exhibits better nuclei segmentation capability compared to the source domain method, especially for missed and overlapping nuclei. In contrast, SIFA\cite{chen2019synergistic} and PDAM\cite{liu2020pdam} exhibit poor segmentation performance on some overlapping nuclei. These results further demonstrate the effectiveness of our proposed method in domain adaptation, whether from pan-cancer to specific cancer types or from pan-cancer to other pan-cancer datasets.

\subsubsection{Detection performance}

\begin{figure}[t]
	\centering
	\includegraphics[width=3.2in]{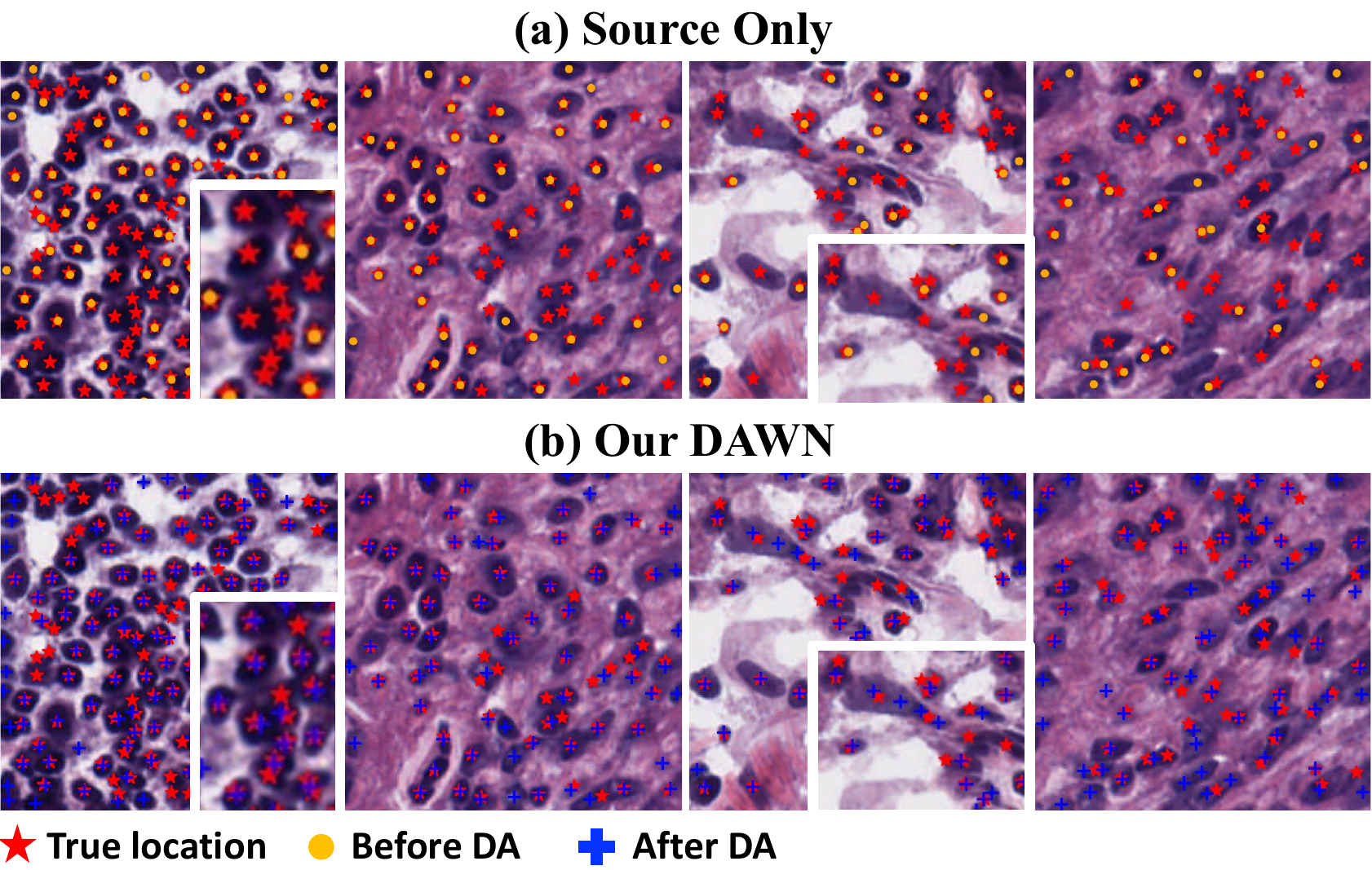}
	\caption{\textbf{The nuclei detection results.} (a) represents the detection results of source only method; (b) represents the detection results of our DAWN.}
	\label{fig:det}
\end{figure}

To evaluate the influence of our proposed auxiliary detection branch on nuclei detection, we conduct a domain adaptation comparison experiment on four target datasets in three metrics, i.e., Recall, Precision, and F1. The experimental results are shown in Table \ref{tab:det}. As seen from the table, when using the detection branch to assist domain transfer, the Recall has been significantly improved without reducing F1, which means that the auxiliary task can help the segmentation network to identify the nucleus more completely. This conclusion can also be verified from the visualization results in Fig. \ref{fig:det}. The figure's red \textcolor{red}{$\star$} represents the actual nuclei location. Orange ``\textcolor{orange}{$\bullet$}" and blue ``\textcolor{blue}{+}" represent the predicted location before and after domain adaptation. The figure shows that after adding the detection task, the predicted results are highly coincident with the actual labels, indicating that our detection task helps enhance the nuclei recognition ability of the segmentation task.

\begin{table}[t]
\centering
\renewcommand{\arraystretch}{1.2}
\caption{The detection results before and after domain adaptation.}
    \setlength{\tabcolsep}{1.2mm}{
        \begin{tabular}{ccccc}
            \hline
            \textbf{Datasets} & \textbf{Stage} & \textbf{Recall} & \textbf{Precision} & \textbf{F1} \\
            \hline
            \textbf{TNBC} & \textit{Bef.} &0.7647 &0.8949  &0.8205	\\
            & \textit{Aft.} &0.8605 (\textbf{0.10$\uparrow$})  & 0.7870 (\textbf{0.11$\downarrow$})  &0.8221 (\textbf{0.002$\uparrow$})	\\
            \textbf{CryoNuSeg} & \textit{Bef.} &0.4637 &0.7522	&0.5737  \\
            & \textit{Aft.} &0.6635 (\textbf{0.20$\uparrow$}) &0.7326 (\textbf{0.02$\downarrow$}) & 0.6963 (\textbf{0.12$\uparrow$})	\\
            \textbf{Lizard} & \textit{Bef.}  &0.7804 &0.8373 &0.7958  \\
            & \textit{Aft.}  &0.8029 (\textbf{0.02$\uparrow$})	&0.8109 (\textbf{0.03$\downarrow$})	&0.8069 (\textbf{0.01$\uparrow$}) \\
            \textbf{ConSeP} & \textit{Bef.} &0.5563 &0.7893 &0.6526	\\
            & \textit{Aft.} &0.7878 (\textbf{0.23$\uparrow$})	&0.5999 (\textbf{0.19$\downarrow$})	&0.6812 (\textbf{0.03$\uparrow$}) \\
            \hline
    \end{tabular}}
    \label{tab:det}
\end{table}

\begin{table}[t]
\centering
\caption{The ablation experiments of CFC and IST modules on TNBC, CryoNuSeg, Lizard and ConSeP datasets. }
\renewcommand{\arraystretch}{1.18}
\resizebox{\linewidth}{!}{
    \begin{tabular}{cccccccc}
        \hline
        \textbf{Data} & \textbf{CFC} & \textbf{IST}& \textbf{DICE} & \textbf{AJI} & \textbf{DQ} & \textbf{SQ} & \textbf{PQ} \\
        \hline
        & & &0.8216	&0.6764	&0.8158	&0.8094	&0.6610	 \\
        \textbf{TNBC} &  & \ding{52} &\underline{0.8405} &\underline{0.7045}	&\underline{0.8346}	&\underline{0.8128}	&\underline{0.6796}	\\
        & \ding{52} & 	&0.8340	&0.7005	&0.8236	&0.8122	&0.6699	 \\
        & \ding{52} & \ding{52} &\textbf{0.8561}	&\textbf{0.7280}	& \textbf{0.8425} & \textbf{0.8202} & \textbf{0.6925}	 \\
        \hline
        & & & 0.7857 & 0.4699 & 0.6121	& \underline{0.7467} & 0.4667 \\
        \textbf{CryoNuSeg} & & \ding{52} & \underline{0.7946} &\underline{0.5006}	&\underline{0.6279}	&0.7404	&\underline{0.4668}\\
        & \ding{52} & &0.7794	&0.4734	&0.5750	&\textbf{0.7479}	&0.4319 \\
        & \ding{52} & \ding{52} &\textbf{0.8041}	&\textbf{0.5079}	&\textbf{0.6374}	&0.7439	&\textbf{0.4760}\\
        \hline
        & & &0.6963	&0.3579	&0.4508	&0.7316	&0.3318 \\
        \textbf{Lizard } & & \ding{52} &\underline{0.7236}	&\textbf{0.4081} &\underline{0.5038} &\underline{0.7344} &\underline{0.3752} \\
        & \ding{52} & & 0.7022 & 0.3627	& 0.4582 & 0.7007	& 0.3476 \\
        & \ding{52} & \ding{52} &\textbf{0.7282}	&\underline{0.4031}	&\textbf{0.5095} &\textbf{0.7364}	&\textbf{0.3768} \\
        \hline
        & &	&0.7813	&0.4859	&0.6101	&\underline{0.7612}	&0.4644 \\
        \textbf{ConSeP} & & \ding{52} &\underline{0.7959}	&\underline{0.5010}	&\underline{0.6146}	&\textbf{0.7621}	&\underline{0.4695}\\
        & \ding{52} & & 0.7899 & 0.4547	& 0.6138 & 0.7584 & 0.4675 \\
        & \ding{52} &\ding{52} &\textbf{0.8047} & \textbf{0.5093} & \textbf{0.6272} & 0.7591 & \textbf{0.4772} \\
        \hline
\end{tabular}}
\label{tab:abl1}
\end{table}

\subsection{Ablation study}

\subsubsection{Evaluation of module designs}

To prove the effectiveness of CFC and IST components, we perform ablation studies with results listed in Table \ref{tab:abl1}. On the TNBC and Lizard datasets, the five metrics have noticeable improvements whether CFC or IST is added. On CryoNuSeg and ConSeP datasets, when the CFC module is used, the DQ and PQ also obtain improvement, and the  DQ and AJI decrease slightly. At the same time, when adding CFC and IST components, the model achieves the best performance, indicating that CFC and IST modules need to be used together. This phenomenon can be attributed to the significant morphological differences between the CryoNuSeg (ConSeP) and the source domain datasets. Only using CFC might lead to feature perturbations of the segmentation network. However, adding IST can guide the network in performing domain transfer toward the target domain. Our ablation study proves that the CFC and IST promote domain adaptation ability and enhance nuclei segmentation performance on weakly supervised scenarios.

To prove that mask fusion and distance filter effectively improve segmentation performance, we perform an ablation study and parameter-sensitive experiments on the TNBC dataset. The results are shown in Table \ref{tab:abl2}. 
From the result, we can see that, whether removing mask fusion or distance filter, the performance decreases evidently, which indicates the two components are effective in the CPL module because they can enhance the quality of pseudo-labels.

\begin{figure}[t]
	\centering
	\includegraphics[width=3.1in]{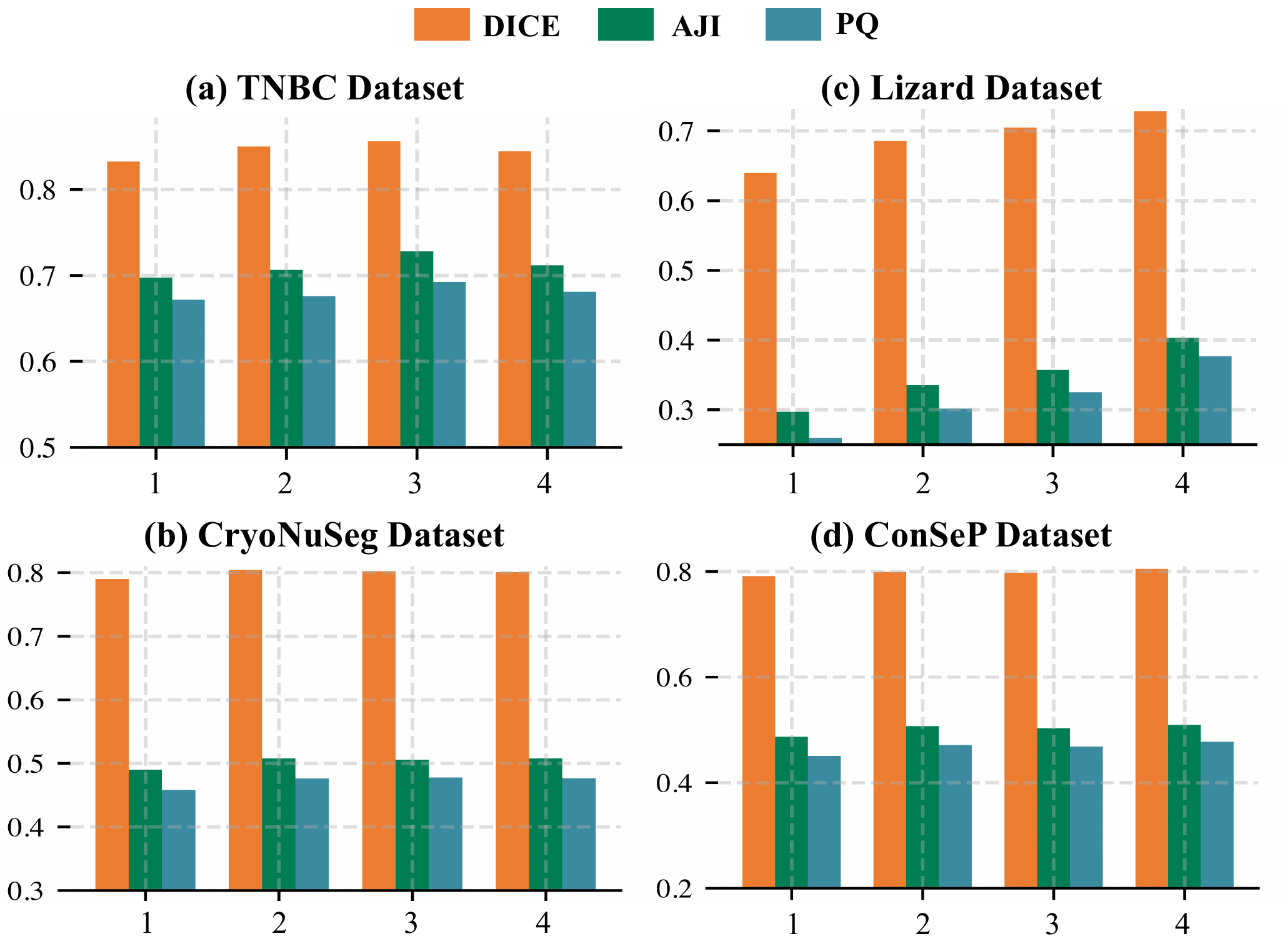}
	\caption{\textbf{The iteration ablation study on four datasets.} The \textit{x}-axis represents the number of iterations, and the \textit{y}-axis represents model performance.}
	\label{fig:iter}
\end{figure}

\begin{table}[t]
	\centering
	\caption{The mask fusion and distance filter ablation experiments on TNBC dataset.}
	\renewcommand{\arraystretch}{1.2}
		\begin{tabular}{c|ccccc}
			\hline
			\multirow{2}{*}{\textbf{Methods}} & \multicolumn{5}{c}{\textbf{PanNuke} $\rightarrow$ \textbf{TNBC}} \\
			\cline{2-6}
			& \textbf{DICE} & \textbf{AJI} & \textbf{DQ} & \textbf{SQ} & \textbf{PQ} \\
			\hline
			w/o fusion & \underline{0.8418} & 0.7116 & \underline{0.8318} & \underline{0.8226} & 0.6845 \\
			w/o distance & 0.8357 & 0.7045	& 0.8258 & \textbf{0.8232} & 0.6814	\\
                ours & \textbf{0.8561} & \textbf{0.7280} & \textbf{0.8425} & 0.8202 & \textbf{0.6925} \\
			\hline
	\end{tabular}
	\label{tab:abl2}
\end{table}

\begin{table}[t!]
	\centering
	\caption{The hyper-parameter $r_1$ ablation experiments on TNBC dataset.}
	\renewcommand{\arraystretch}{1.2}
		\begin{tabular}{c|ccccc}
			\hline
			\multirow{2}{*}{\textbf{Methods}} & \multicolumn{5}{c}{\textbf{PanNuke} $\rightarrow$ \textbf{TNBC}} \\
			\cline{2-6}
			& \textbf{DICE} & \textbf{AJI} & \textbf{DQ} & \textbf{SQ} & \textbf{PQ} \\
			\hline
			$r_1$=5 & 0.8147 & 0.6828 & 0.7914 & 0.7784 & 0.6554 \\
                $r_1$=11 & \textbf{0.8561} & \textbf{0.7280} & \textbf{0.8425} & \textbf{0.8202} & 0.6925 \\
			$r_1$=20 & 0.8323 & 0.7024 & 0.8191	& 0.7923 & \textbf{0.7022} \\
			\hline
	\end{tabular}
	\label{tab:para1}
\end{table}

\subsubsection{Evaluation of Hyper-parameter}

First, by observing Table \ref{tab:para1}, there is a significant decrease in the performance of the model when the radius range of the nuclei is altered. Whether $r_1$ is set to $5$ or $20$, the Dice coefficient drops by more than 2 \%. This indicates that the setting of nucleus radius has a substantial impact on the model and that configuring parameters based on the radius from the validation set is advantageous. Secondly, observing Tables \ref{tab:para2} and \ref{tab:para3} it can be seen that there is no significant variation in model performance when modifying the standard deviation $\sigma$ and $\theta$ of the nuclear radius. In fact, in some metrics, the model even outperforms the reported results in the main text; for example, when $\sigma$ = 4, the SQ metric improves by 0.2\%. However, this variation is not substantial, indicating that the model parameters are less affected by $\sigma$ and $\theta$. Lastly, we also include additional sensitivity analyses of distance $d$ as presented in Table \ref{tab:para4}.

\subsubsection{Evaluation of Iterative Training}

\begin{figure}[t!]
	\centering
	\includegraphics[width=3.0in]{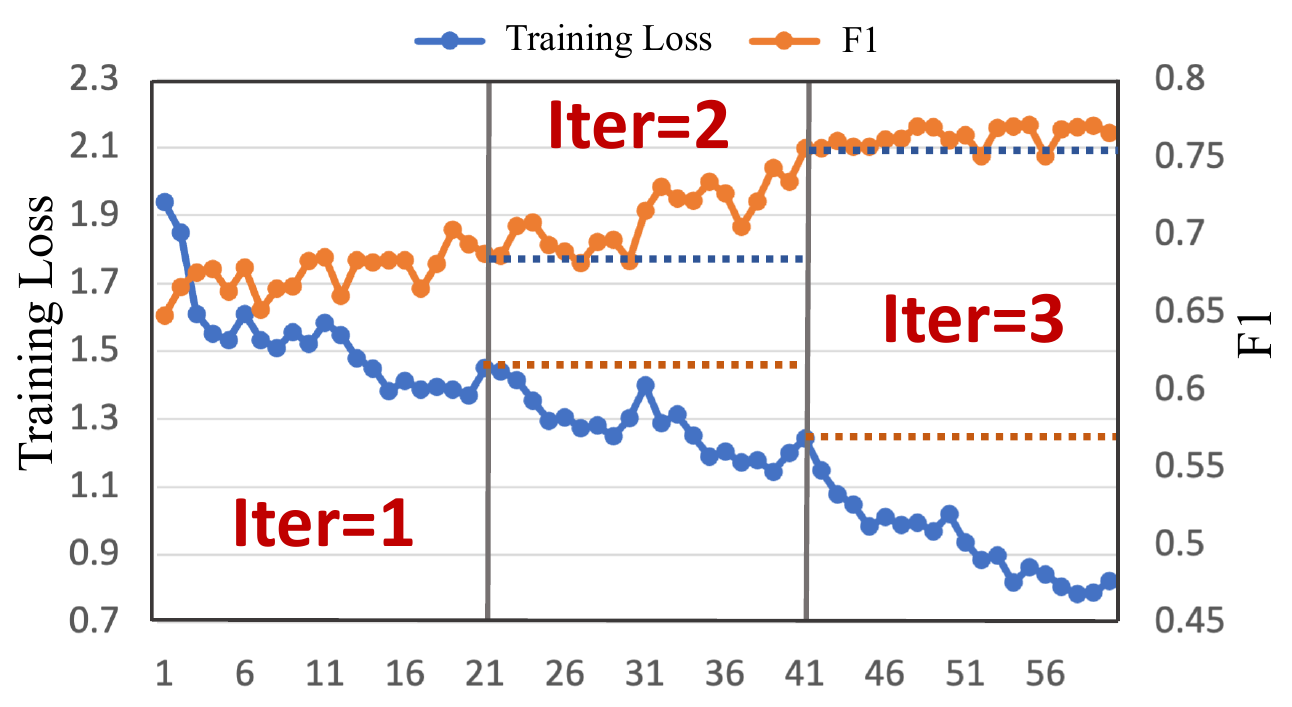}
	\caption{\textbf{The iteration training loss on TNBC dataset.} The \textit{x}-axis represents the training step and the \textit{y}-axis represent the training loss and F1 value.}
	\label{fig:loss}
\end{figure}

We perform the experiments across the four datasets to evaluate the impact of the number of iterations in Fig. \ref{fig:iter}. In this figure, the \textit{x}-axis represents the number of iterations, corresponding to the parameter "\textit{I}" in Algorithm \ref{algorithm1}, while the \textit{y}-axis represents the values of the DICE, AJI, and PQ metrics. 
The results show that on the TNBC and Lizard datasets, the model's performance shows an increasing trend as the number of training iterations increases. Specifically, the TNBC dataset achieves optimal performance at \textit{I}=3, while the Lizard dataset performs best at \textit{I}=4. In contrast, on the CryoNuSeg and ConSeP datasets, the model's performance remains relatively stable after the second round of training. Consequently, we select the optimal performance according to \textit{I} as the segmentation result. Furthermore, Fig. \ref{fig:loss} presents the loss function and the F1 curve for three training iterations in the TNBC dataset, each iteration consisting of 20 training epochs. From this figure, it can be observed that as training progresses, the model's loss steadily decreases. Correspondingly, the F1 score increases with training and performs best in the final iteration.

\begin{table}[t!]
	\centering
	\caption{The hyper-parameter $\sigma$ ablation experiments on TNBC dataset.}
	\renewcommand{\arraystretch}{1.2}
		\begin{tabular}{c|ccccc}
			\hline
			\multirow{2}{*}{\textbf{Methods}} & \multicolumn{5}{c}{\textbf{PanNuke} $\rightarrow$ \textbf{TNBC}} \\
			\cline{2-6}
			& \textbf{DICE} & \textbf{AJI} & \textbf{DQ} & \textbf{SQ} & \textbf{PQ} \\
			\hline
			$\sigma$=2.0 & 0.8447 & 0.7120	& 0.8394 & 0.8191 & 0.7022 \\
                $\sigma$=2.75 & \textbf{0.8561} & \textbf{0.7280} & \textbf{0.8425} & 0.8202 & \textbf{0.6925}	 \\
			$\sigma$=4.0 & 0.8402 & 0.7221	& 0.8334 & \textbf{0.8227} & 0.6896 \\
			\hline
	\end{tabular}
	\label{tab:para2}
\end{table}

\begin{table}[t!]
	\centering
	\caption{The hyper-parameter $\theta$ ablation experiments on TNBC dataset.}
	\renewcommand{\arraystretch}{1.2}
		\begin{tabular}{c|ccccc}
			\hline
			\multirow{2}{*}{\textbf{Methods}} & \multicolumn{5}{c}{\textbf{PanNuke} $\rightarrow$ \textbf{TNBC}} \\
			\cline{2-6}
			& \textbf{DICE} & \textbf{AJI} & \textbf{DQ} & \textbf{SQ} & \textbf{PQ} \\
			\hline
			$\theta$=0.1 & 0.8489 & 0.7233	& \textbf{0.8510} & 0.8212 & 0.6917 \\
                $\theta$=0.2 & \textbf{0.8561} & \textbf{0.7280} & 0.8425 & 0.8202 & \textbf{0.6925}	 \\
			$\theta$=0.3 & 0.8550 & 0.7226 & 0.8390 & \textbf{0.8226} & 0.6898 \\
			\hline
	\end{tabular}
	\label{tab:para3}
\end{table}

\begin{table}[t]
	\centering
	\caption{The hyper-parameter $d$ ablation experiments on TNBC dataset.}
	\renewcommand{\arraystretch}{1.2}
		\begin{tabular}{c|ccccc}
			\hline
			\multirow{2}{*}{\textbf{Methods}} & \multicolumn{5}{c}{\textbf{PanNuke} $\rightarrow$ \textbf{TNBC}} \\
			\cline{2-6}
			& \textbf{DICE} & \textbf{AJI} & \textbf{DQ} & \textbf{SQ} & \textbf{PQ} \\
			\hline
			$d$=10 & 0.6541	& 0.4917 & 0.5771	& 0.7054 & 0.4170 \\
                $d$=25 & \textbf{0.8561} & \textbf{0.7280} & \textbf{0.8425} & 0.8202 & \textbf{0.6925}	 \\
			$d$=40 & 0.8332 & 0.7120 & 0.8313 & \textbf{0.8220} & 0.6850 \\
			\hline
	\end{tabular}
	\label{tab:para4}
\end{table}

\begin{figure}[t!]
	\centering
	\includegraphics[width=3.5in]{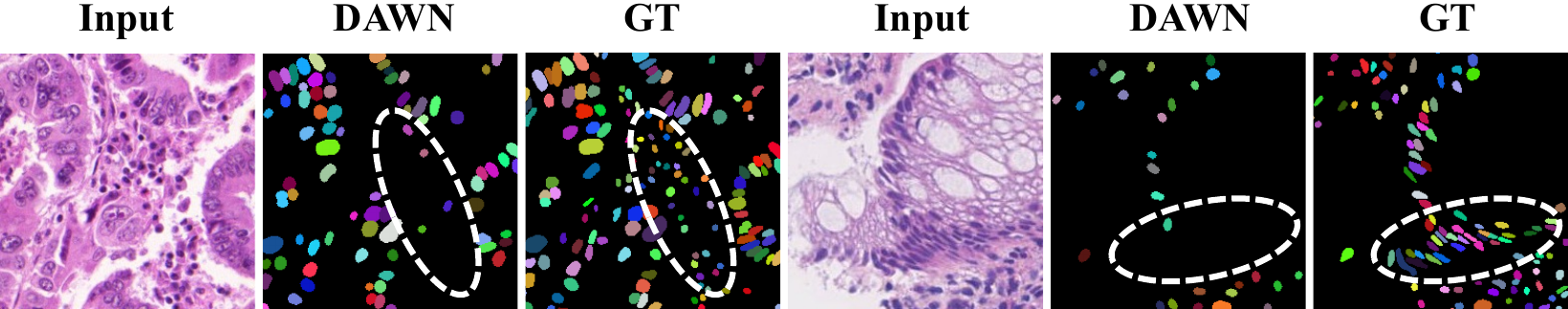}
	\caption{The limitation of our proposed DAWN when facing nuclei with significant morphological variations.}
	\label{fig:future}
\end{figure}

\section{Conclusion and Limitation}

In this paper, to reduce the reliance on manual annotations for nuclei segmentation, we propose a domain-adaptive weakly supervised segmentation method, termed DAWN. Our approach enables the transfer of a model trained on public data (source domain) to arbitrary external datasets (target domain) without requiring fine pixel-level annotations. This domain adaptation is guided by an auxiliary detection task. DAWN employs a Consistent Feature Constraint (CFC) module to improve domain transfer efficiency at the feature level. Additionally, we introduce a combined pseudo-label optimization method (CPL) and an interactive supervision training (IST) strategy to enhance domain adaptation at the task level. Ablation experiments confirm the effectiveness of these components in improving nuclei segmentation performance.

Despite these advancements, DAWN exhibits limitations when handling nuclei with significant intra-image morphological variations, such as large spindle-shaped nuclei coexisting with small, circular ones, as illustrated in Fig. \ref{fig:future}. This leads to occasional segmentation errors, particularly under-segmentation of nuclei with atypical shapes. To address these challenges, future work will focus on developing enhanced interaction mechanisms between the segmentation and detection networks. Specifically, we aim to introduce a multi-task learning framework that simultaneously optimizes segmentation and shape detection, allowing the model to better differentiate and capture varying nuclear shapes by leveraging contextual and spatial information across scales. Additionally, we plan to incorporate a region-aware feature aggregation module, which will allow the model to adaptively focus on regions with high morphological diversity. This approach aims to improve segmentation accuracy for nuclei with atypical geometries by explicitly modeling the relationships between various shapes within the same image. By combining these stratesgies, we aim to improve DAWN’s ability to segment diverse and complex nuclei, pushing the boundaries of domain adaptation in histopathology image analysis.

\bibliographystyle{IEEEtran}
\bibliography{egbib}

\begin{IEEEbiography}[{\includegraphics[width=1.0in,height=1.25in,clip,keepaspectratio]{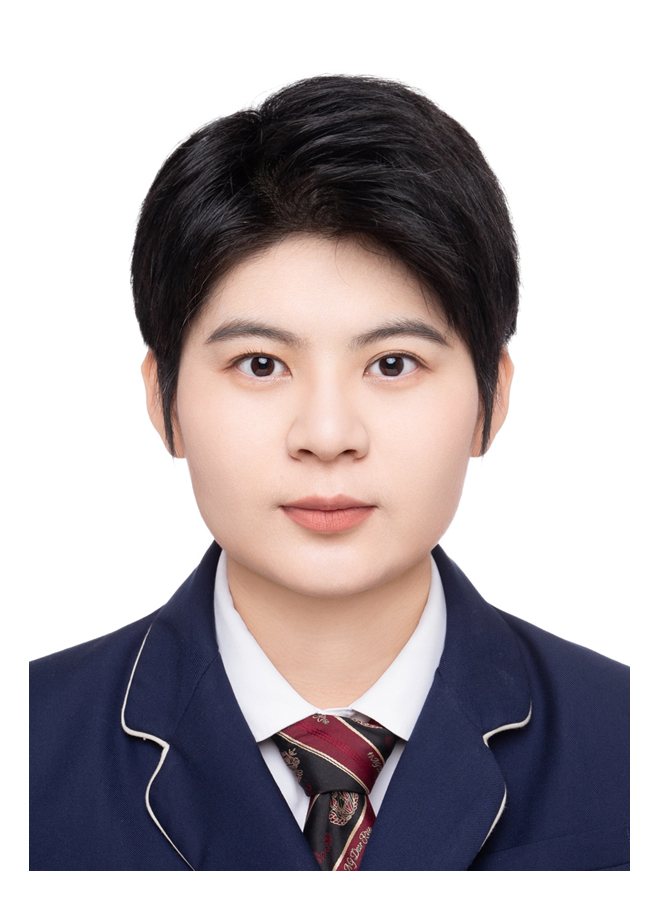}}]{Ye Zhang} received the BS degree from the School of Economics and Management, Beijing Forestry University, Beijing, China, in 2017, and the MS degree from the School of Mathematics and Statistics, Lanzhou University, Lanzhou, China, in 2021. She is currently pursuing the PhD degree at the Harbin Institute of Technology, Shenzhen, China, and is a joint PhD student at Leibniz-Institut für Analytische Wissenschaften – ISAS – e.V., Germany. Her research interests include biomedical image analysis and instance segmentation.
\end{IEEEbiography}

\begin{IEEEbiography}[{\includegraphics[width=1.0in,height=1.25in,clip,keepaspectratio]{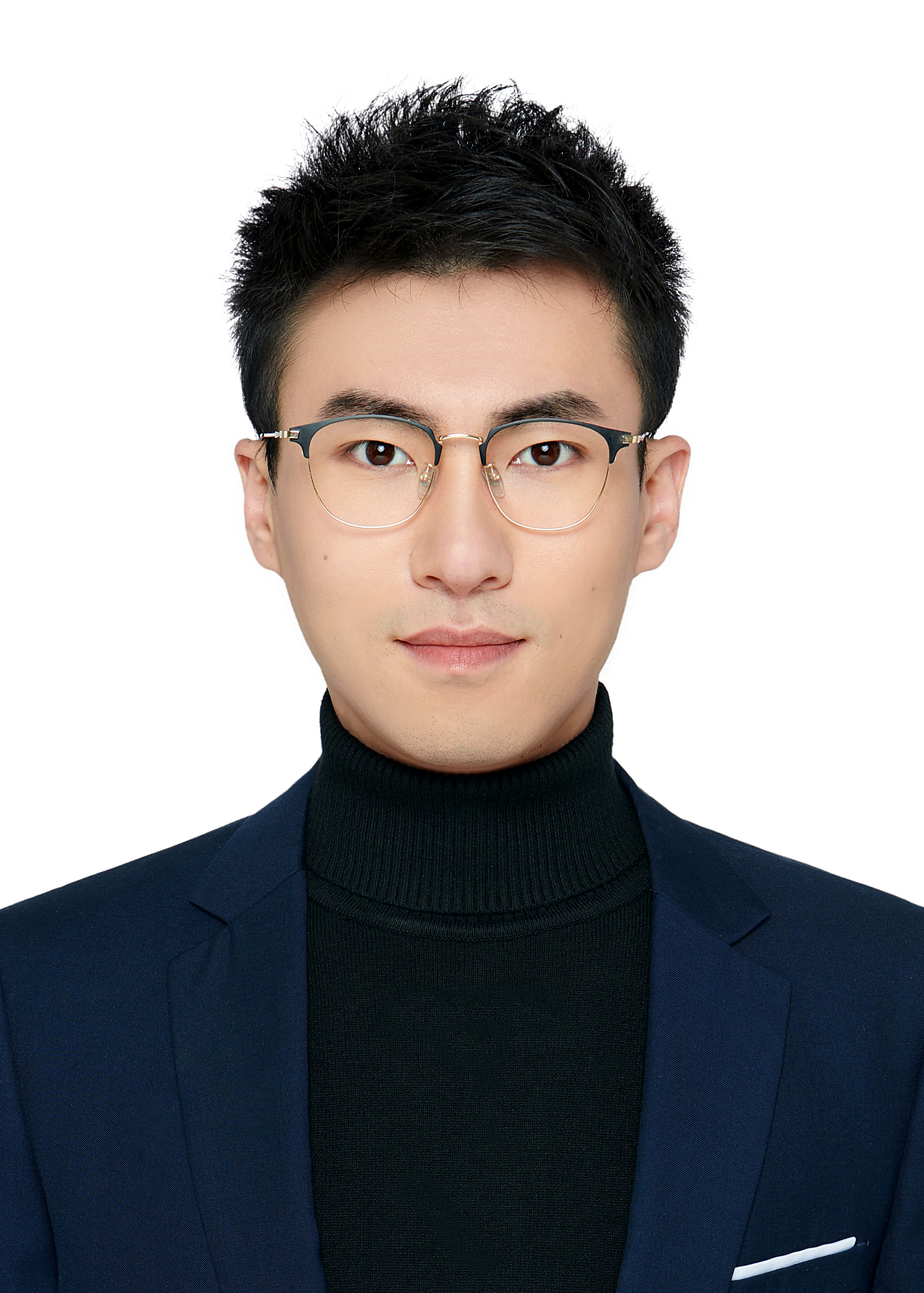}}]{Yifeng Wang} received the MS degree in Harbin Institute of Technology Shenzhen Graduate School, where he is currently working toward the PhD degree. His research interests include machine learning model design and analysis, signal processing, human activity recognition, and computer vision.
\end{IEEEbiography}

\begin{IEEEbiography}[{\includegraphics[width=1.0in,height=1.25in,clip,keepaspectratio]{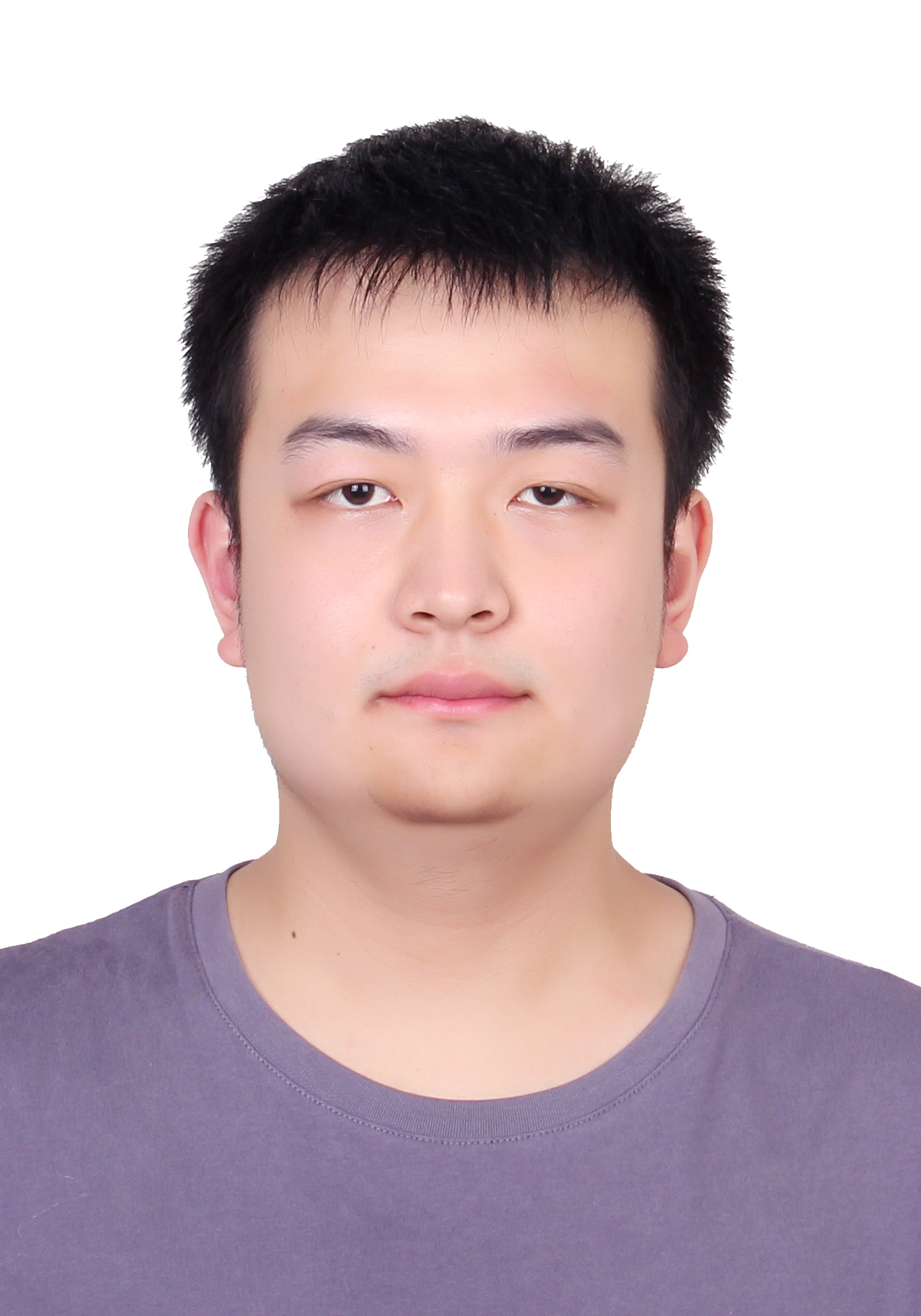}}]{Zijie Fang} received the BS degree in Software Engineering from Nanjing University of Information Science and Technology in 2022. He is currently working towards the MS degree at Tsinghua Shenzhen International Graduate School, Tsinghua University. His research interests include computational pathology and medical image segmentation.
\end{IEEEbiography}

\begin{IEEEbiography}[{\includegraphics[width=1.0in,height=1.25in,clip,keepaspectratio]{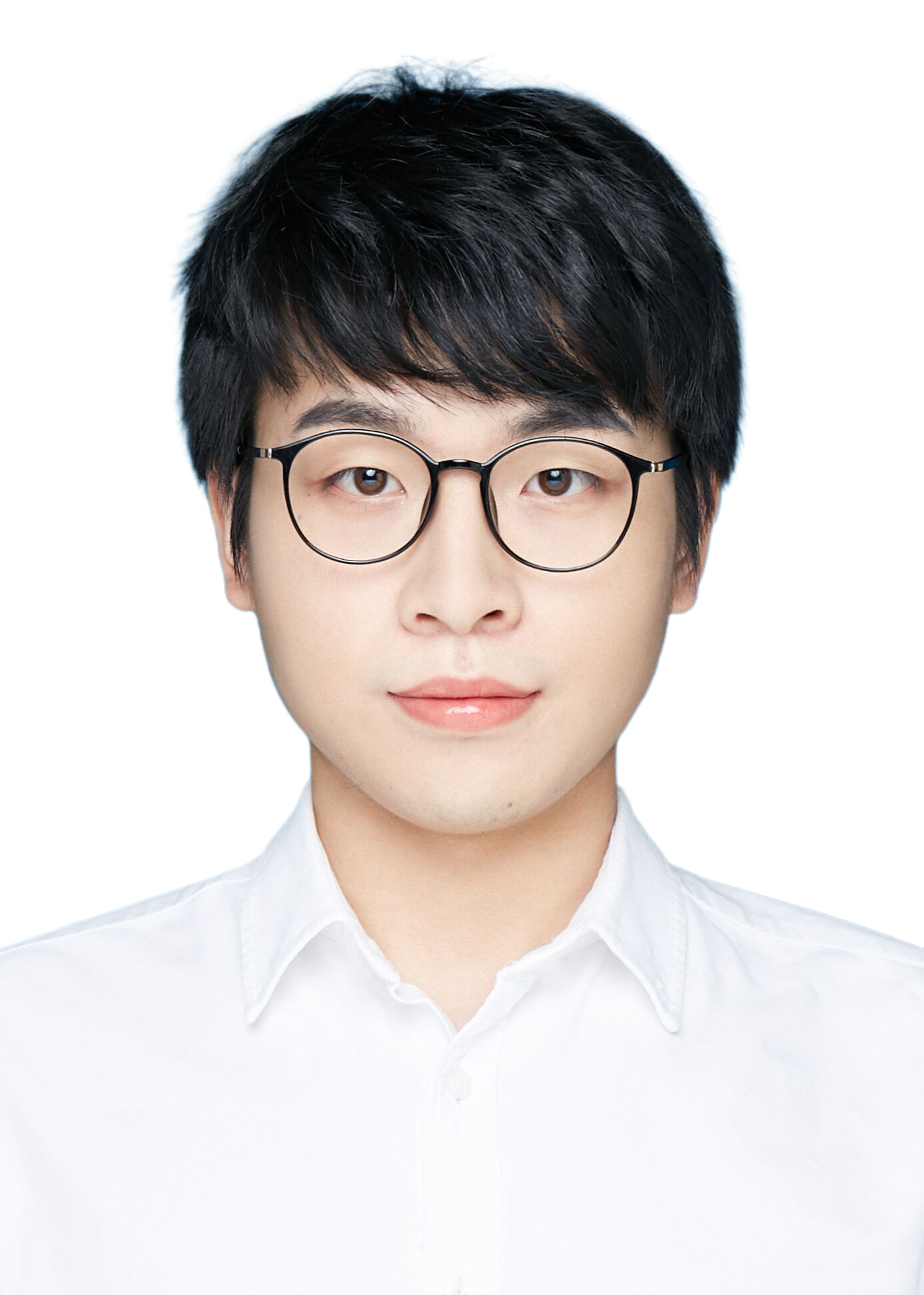}}]{Hao Bian} received the BS degree in Engineering from Hefei University of Technology in 2021. Currently, he is pursuing a MS degree at Tsinghua University. His research interests span across the fields of computer vision, low-level vision, and medical image analysis.
\end{IEEEbiography}

\begin{IEEEbiography}[{\includegraphics[width=1.0in,height=1.25in,clip,keepaspectratio]{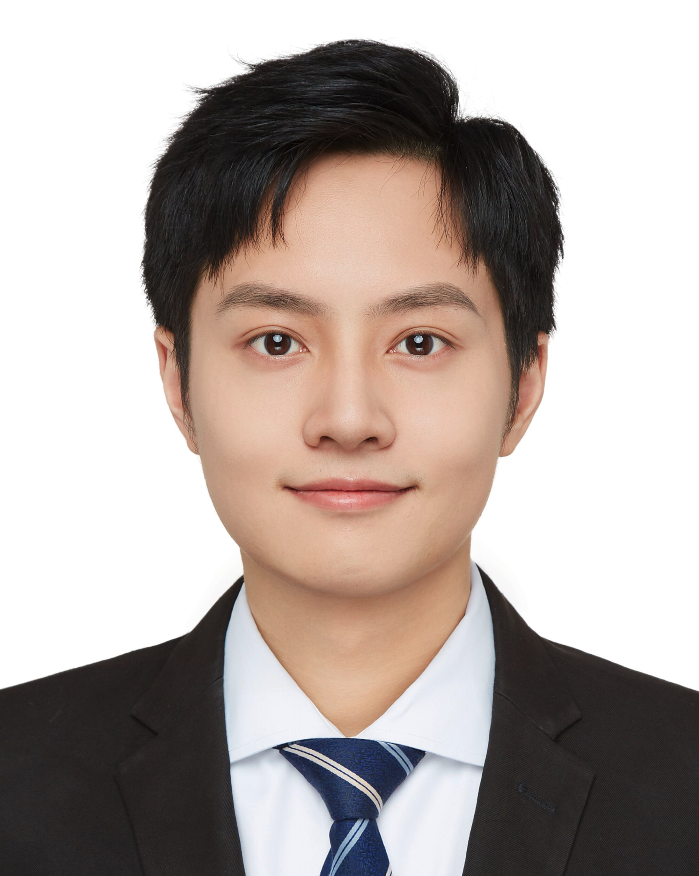}}]{Linghan Cai} received the BS degreee from the Department of Information and Electrical Engineering, China Agricultural University, in 2020, the M.S. degree in the Department of Electronic Information Engineering, Beihang University, in 2023. He is currently pursuing the PhD degree with Harbin Institute of Technology. His research interests include scene parsing and multi-modal learning.
\end{IEEEbiography}

\begin{IEEEbiography}[{\includegraphics[width=1.0in,height=1.25in,clip,keepaspectratio]{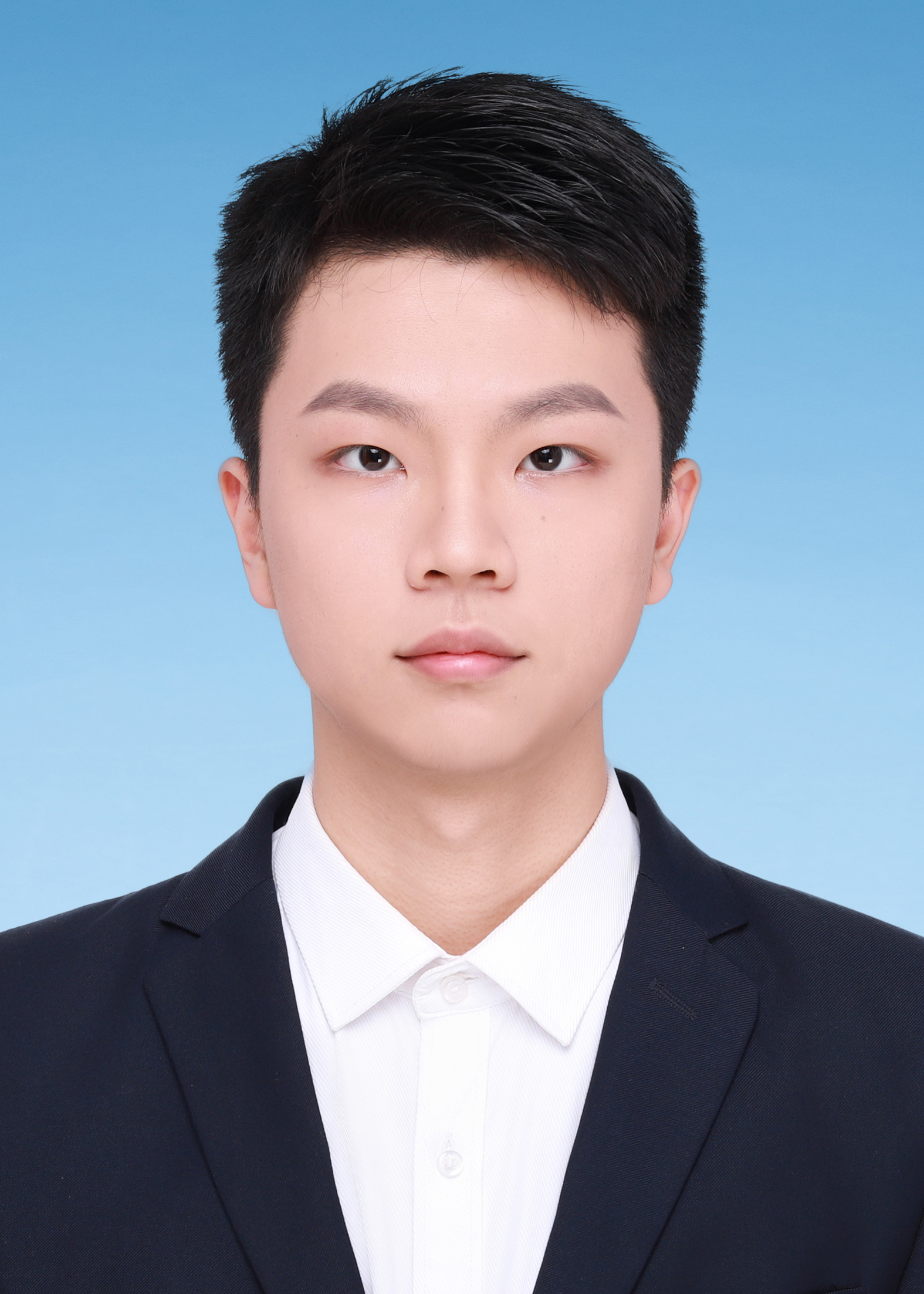}}]{Ziyue Wang} receives the BS degree from the Department of Computer Science and Technology,  Harbin Institute of Technology (Shenzhen) in 2024, and he is working towards the MS degree at the School of Electrical and Computer Engineering, National University of Singapore. His research interests include medical image analysis and foundation models in computer vision.
\end{IEEEbiography}

\begin{IEEEbiography}[{\includegraphics[width=1.0in,height=1.25in,clip,keepaspectratio]{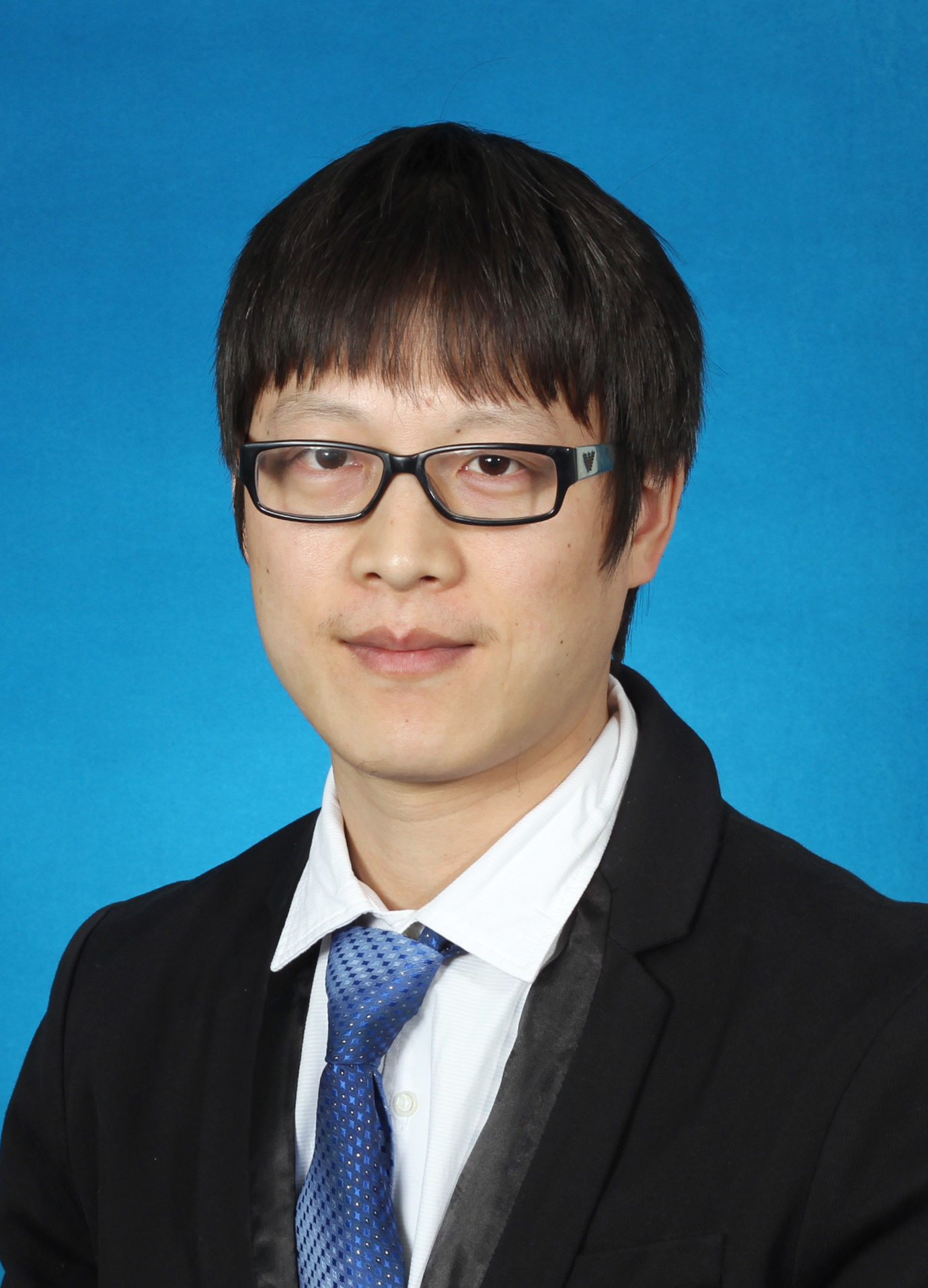}}]{Yongbing Zhang} received the BA degree in English and the MS and PhD degrees in computer science from Harbin Institute of Technology, Harbin, China, in 2004, 2006, and 2010, respectively. He is currently a Professor of Computer Science and Technology,Harbin Institute of Technology (Shenzhen), Shenzhen, China. He joined Tsinghua Shenzhen International Graduate School from 2010 to 2020, and was a visiting scholar of University of California, Berkeley from 2016 to 2017. He was the receipt of the Best Student Paper Award at IEEE International Conference on Visual Communication and Image Processing in 2015 and the Best Paper Award at Pacific-Rim Conference on Multimedia in 2018. His current research interests include signal processing, machine learning, and computational imaging.
\end{IEEEbiography}

\end{document}